\documentclass{aa}
\newif\iffigure
\figurefalse
\figuretrue

\newcommand{\AK}[1]{\color{black}#1}
\newcommand{\rev}[1]{\color{black}#1}
\newcommand{\revsec}[1]{\color{black}#1}
\newcommand{\revtir}[1]{\color{black}#1}
\newcommand{\revfor}[1]{\color{black}#1}
\newcommand{\revfif}[1]{\color{black}#1}
\newcommand{\revsix}[1]{\color{black}#1}
\newcommand{\editage}[1]{\color{black}#1}
\newcommand{\revsev}[1]{\color{black}#1}

\newcommand{\reveig}[1]{\color{black}#1}
\newcommand{\revnin}[1]{\color{black}#1}
\newcommand{\revten}[1]{\color{black}#1}
\newcommand{\revele}[1]{\color{black}#1}

\usepackage{graphicx,natbib,url,twoopt}
\usepackage[varg]{txfonts}           
\usepackage{hyperref}              
\usepackage{pdfcomment}              
\usepackage{acronym}                 
\usepackage{comment}
\usepackage{cases}
\usepackage{empheq}                  
\usepackage{here}
\usepackage{bm}
\usepackage[version=3]{mhchem}
\hypersetup{
 colorlinks=true,%
 linkcolor=blue,
 citecolor=blue,
}

\usepackage{lineno}
\newcommand*\patchAmsMathEnvironmentForLineno[1]{
  \expandafter\let\csname old#1\expandafter\endcsname\csname #1\endcsname
  \expandafter\let\csname oldend#1\expandafter\endcsname\csname end#1\endcsname
  \renewenvironment{#1}
     {\linenomath\csname old#1\endcsname}
     {\csname oldend#1\endcsname\endlinenomath}}
\newcommand*\patchBothAmsMathEnvironmentsForLineno[1]{
  \patchAmsMathEnvironmentForLineno{#1}
  \patchAmsMathEnvironmentForLineno{#1*}}
\AtBeginDocument{
\patchBothAmsMathEnvironmentsForLineno{equation}
\patchBothAmsMathEnvironmentsForLineno{align}
\patchBothAmsMathEnvironmentsForLineno{flalign}
\patchBothAmsMathEnvironmentsForLineno{alignat}
\patchBothAmsMathEnvironmentsForLineno{gather}
\patchBothAmsMathEnvironmentsForLineno{multline}
}

\bibpunct{(}{)}{;}{a}{}{,}    

\makeatletter
\newcommand{\bibnote}[2]{\global\@namedef{#1note}{#2}}
\newcommand{\biblink}[2]{\global\@namedef{#1link}{#2}}
\newcommand{\Tabref}[1]{Table~\ref{#1}}
\newcommand{\Equref}[1]{Eq.~(\ref{#1})}
\newcommand{\Figref}[1]{Fig.~\ref{#1}}
\makeatother

\makeatletter
 \newcommandtwoopt{\citeads}[3][][]{%
   \nonstopmode
   \href{http://adsabs.harvard.edu/abs/#3}%
        {\def\hyper@linkstart##1##2{}%
         \let\hyper@linkend\@empty\citealp[#1][#2]{#3}}
   \biblink{#3}{\href{http://adsabs.harvard.edu/abs/#3}{ADS}}%
   \errorstopmode}            
 \newcommandtwoopt{\citepads}[3][][]{%
   \nonstopmode
   \href{http://adsabs.harvard.edu/abs/#3}%
        {\def\hyper@linkstart##1##2{}%
         \let\hyper@linkend\@empty\citep[#1][#2]{#3}}
   \biblink{#3}{\href{http://adsabs.harvard.edu/abs/#3}{ADS}}%
   \errorstopmode}            
 \newcommandtwoopt{\citetads}[3][][]{%
   \nonstopmode
   \href{http://adsabs.harvard.edu/abs/#3}%
        {\def\hyper@linkstart##1##2{}%
         \let\hyper@linkend\@empty\citet[#1][#2]{#3}}
   \biblink{#3}{\href{http://adsabs.harvard.edu/abs/#3}{ADS}}%
   \errorstopmode}            
 \newcommandtwoopt{\citeyearads}[3][][]{%
   \nonstopmode
   \href{http://adsabs.harvard.edu/abs/#3}%
        {\def\hyper@linkstart##1##2{}%
         \let\hyper@linkend\@empty\citeyear[#1][#2]{#3}}
   \biblink{#3}{\href{http://adsabs.harvard.edu/abs/#3}{ADS}}%
   \errorstopmode}            
\makeatother

\newacro{ADS}{Astrophysics Data System}
\newacro{NLTE}{non-local thermodynamic equilibrium}
\newacro{NASA}{National Aeronautics and Space Administration}

\begin{document}
\authorrunning{A. Kuwahara et al.}
\titlerunning{Dust ring and gap formation by gas flow}
   \title{Dust ring and gap formation by gas flow induced by low-mass planets \revsec{embedded in protoplanetary disks}}
   \subtitle{$\rm I$. Steady\revsec{-}state model}
      \author{Ayumu Kuwahara\inst{1,2}
          \thanks{\email{kuwahara@elsi.jp}} 
          \and Hiroyuki Kurokawa\inst{2}
          \and Takayuki Tanigawa\inst{3}
          \and Shigeru Ida\inst{2}}

   \institute{Department of Earth and Planetary Sciences, Tokyo Institute of Technology, Ookayama, Meguro-ku, Tokyo, 152-8551, Japan
         \and
             Earth-Life Science Institute, Tokyo Institute of Technology, Ookayama, Meguro-ku, Tokyo, 152-8550, Japan
         \and
             National Institute of Technology, Ichinoseki College, Takanashi, Hagisho, Ichinoseki-shi, Iwate 021-8511, Japan}

   \date{Received September XXX; accepted YYY}

 
  \abstract
   {\rev{Recent high-spatial-resolution observations have revealed dust substructures in protoplanetary disks such as rings and gaps\editage{,} \revfif{which do not always correlate with gas}.} \revtir{Because radial gas flow induced by low-mass, non-gas-gap-opening planets \revfif{could affect} the radial drift of dust, it potentially forms \revsix{these} dust substructures in disks.}}
   {\rev{We \revsec{investigate} the potential of gas flow induced by low-mass planets to sculpt the rings and gaps in the dust profiles.}}
   {\rev{We first perform three-dimensional hydrodynamical simulations, which resolve the local \revtir{gas} flow past a planet. \revsec{We then calculate the trajectories of dust influenced by the planet-induced gas flow. Finally, we compute the steady-state dust surface density by incorporating the influences of the planet-induced gas flow into \editage{a} one-dimensional \revfif{dust advection-diffusion model}.}}}
   {\revsec{The outflow of the gas toward \editage{the outside of} the planetary orbit inhibits the radial drift of dust, \editage{leading} to dust accumulation (the dust ring). The outflow toward \editage{the inside of} the planetary orbit enhances the inward drift of dust, causing dust depletion around the planetary orbit (the dust gap).} \rev{Under weak turbulence ($\alpha_{\rm diff}\lesssim10^{-4}$, where $\alpha_{\rm diff}$ is the turbulence strength \revsec{parameter}), the gas flow induced by the planet with $\gtrsim1\,M_{\oplus}$ (Earth mass) generates the dust ring and gap \revtir{in the distribution of small dust grains ($\lesssim1$ cm)} with the radial \revtir{extent} of $\sim1\text{--}10$ times gas scale height around the planetary orbit \revsec{without creating a gas gap and pressure bump.}}}
  {\revtir{The gas flow induced by low-mass, non-gas-gap-opening planets can be considered a possible origin of the observed dust substructures in disks. Our results may be helpful to \revfif{explain the disks whose dust substructures were found not to correlate with those of the gas.}}}
  
  
    \keywords{Hydrodynamics --
                Planets and satellites: formation --
                Protoplanetary disks}

   \maketitle

%

\section{Introduction}\label{sec:Introduction}
Recent high-\AK{spatial}-resolution observations have revealed \rev{substructures in the dust profiles (hereafter dust substructures)\editage{,} such as rings and gaps} in \AK{both young and evolved} protoplanetary disks \citep[e.g.,][]{ALMA:2015,andrews2018disk,long2018gaps,long2020dual,van2019protoplanetary}. Several mechanisms\editage{---planet-free and planet-dependent---}have been proposed to explain the dust rings and gaps in disks. 

\rev{The planet-free mechanisms} include dust accumulation at the edge of the dead-zone \citep{flock2015gaps}, dust growth at snow lines \citep{zhang2015evidence}, \AK{dust sintering \citep{okuzumi2016sintering}}, secular gravitational instability \citep[SGI;][]{Takahashi:2014,Takahashi:2016,tominaga2018non,tominaga2019revised,tominaga2020secular}, self-induced dust pile-up \citep{gonzalez2015alma,gonzalez2017self}, magnetic disk winds \citep{suriano2017rings,suriano2018formation}, self-organization process due to nonideal magnetohydrodynamical (MHD) effects \citep{bethune2017global}, hypothetical axisymmetric radial pressure bump \citep{dullemond2018disk}, \rev{and} thermal wave instability \citep{watanabe2008thermal,ueda2019dust,ueda2021thermal,wu2021irradiation}. 


\reveig{Disk-planet interaction is the other mechanism to interpret the dust rings and gaps in disks. \cite{paardekooper2004planets} \revnin{was} the first to discover that planets can open dust gaps in gas disks.} A planet embedded in a disk perturbs the surrounding gas. The tidal torque by a giant planet can remove gas from the vicinity of the planetary orbit \citep{Goldreich:1979,Goldreich:1980,lin1979tidal}\editage{, leading to the formation of} a gas gap in a disk \citep[e.g.,][]{lin1986tidal,kley1999mass,nelson2000migration,crida2006width,duffell2013gap,fung2014empty,kanagawa2015mass}. The young protoplanetary disks contain abundant millimeter-centimeter (mm--cm)-sized \rev{dust} particles \cite[e.g.,][]{testi2003large}. These \rev{particles} drift from the outer to the inner region of the disk due to the global pressure
gradient \cite[\AK{the radial drift of dust;}][]{Whipple:1972,Weidenschilling:1977a}. \editage{Dust} \rev{particles} can be trapped at the edge of the gas gap, \revsec{which is} \AK{the so-called dust filtration mechanism}, \editage{forming} \revsec{a dust ring and a gap} \citep[e.g.,][]{rice2006dust,paardekooper2006dust,pinilla2012ring,dong2015observational,kanagawa2018impacts}. \rev{Thus, \revsec{in the case of the dust substructure created by a gas gap}, one would expect the spatial distribution of dust substructures \editage{to be} correlated with that of gas.}

\AK{However,} \revsec{the spatial distribution of dust does not always correlate with that of gas} \citep[e.g.,][]{zhang2021molecules,jiang2021no}. \rev{This poses a \editage{challenge for the dust substructure formation model predictions} by \rev{gas-gap-opening} planet\revsec{s for several systems}.} \AK{Moreover,} rings and gaps seem to be ubiquitous \AK{in the outer region of large, bright protoplanetary disks \citep[$\gtrsim10$ au; e.g., ][]{andrews2018disk}}, but the occurrence rate of giant planets \AK{per star is limited} to $\sim10$\% \citep{Johnson:2010,Fernandes:2019}. \revsec{The period-mass distribution of putative giant planets inferred from the observed dust gap widths is not reconciled with the current observed period-mass distribution of giant planets \citep{bae2018diverse,lodato2019newborn,diskdynamics2020visualizing}.} \revtir{Although this discrepancy might be solved by considering large-scale migration and planetary growth} \citep{lodato2019newborn}, it is not conclusive that \rev{\editage{most} of dust substructures} are derived from interactions between disk\revsec{s} and \rev{gas-gap-opening} planet\revsec{s}. 

\revsec{The ambiguous relationship between the spatial distribution of gas and dust may be solved by considering low-mass planets that do not carve a deep gas gap in a disk.} \rev{Here\editage{,} we focus on the potential of gas flow induced by low-mass, non-\rev{gas-gap-opening} planets to sculpt dust substructures in protoplanetary disks. Recent hydrodynamical simulations have revealed that a low-mass planet embedded in a disk induces gas flow with a complex three-dimensional (3D) structure \citep{Ormel:2015b,Fung:2015,Fung:2019,Lambrechts:2017,Cimerman:2017,Kurokawa:2018,Kuwahara:2019,Bethune:2019,moldenhauer2021steady}. The key point is the outflow of the gas, which occurs in the radial direction \editage{to} the disk. The outflow of the gas deflects \editage{the trajectories of dust particles significantly}, even \editage{in the case of low-mass planets} \citep[$\sim1\,M_{\oplus}$;][]{Kuwahara:2020a,Kuwahara:2020b}. Once \editage{the radial outward gas flow disrupts the motion of the particles}, the inward drift of particles could be \editage{inhibited, leading to the formation of a dust substructure. This mechanism is studied in this paper.}}
Since planet-induced gas flow is a local phenomenon, a high-resolution simulation is \editage{required} to capture the detailed 3D structure of the gas flow within the \revsec{Bondi and Hill spheres}. However, the \editage{the influence of the} 3D planet-induced gas flow is missing in previous \revsec{studies on the effects of planets on the global dust distribution in disks.} \rev{So far, the influence of the planet-induced gas flow on the dust particles has been investigated in terms of the accretion rate of particles onto the growing planet \citep{Popovas:2018a,Popovas:2018b,Kuwahara:2020a,Kuwahara:2020b,okamura2021growth}}.

The structure of this paper is as follows. In Sect. \ref{sec:methods} we describe \revsec{our} numerical method\revsec{s}. In Sect. \ref{sec:result} we show the results obtained from a series of numerical calculations. In Sect. \ref{sec:discussion} we discuss the implications \revtir{planet formation theory and observations of protoplanetary disks}. We \rev{conclude} in Sect. \ref{sec:conclusion}. \revfif{The readers who are not interested in the details of modeling and physical aspects of \editage{the} results can skip Sects. \ref{sec:Three-dimensional hydrodynamical simulations (the local domain)}--\ref{sec:Width of forbidden region} and \editage{proceed} directly to Sect. \ref{sec:Influence of planet-induced gas flow on the dust surface density} and \editage{beyond}.}

\section{Methods}\label{sec:methods}
\subsection{Model overview}
\rev{In this section, we provide an overview of our model. Our approach is summarized in \Figref{fig:config}.} \revtir{First,} \rev{assuming a \reveig{compressible}, inviscid sub-Keplerian gas disk \revten{with the vertical stratification due to the vertical component of the stellar gravity}, we performed 3D \reveig{nonisothermal} hydrodynamical simulations. \reveig{Radiative cooling is implemented by using the $\beta$ cooling model \citep{Gammie:2001}.} \revnin{The self-gravity of gas is neglected.} \editage{Hydrodynamical} simulations were performed in the frame co-rotating with the planet (the black sphere in \Figref{fig:config}). \revsec{\editage{These} simulations} resolve the \editage{region's local gas flow, including the} planet's Hill and Bondi spheres (Sect. \ref{sec:Three-dimensional hydrodynamical simulations (the local domain)}).}

\rev{Second, we calculated the 3D trajectories of dust particles influenced by the planet-induced gas flow (Sect. \ref{sec:Three-dimensional-orbital-calculation-of-dust-particles}). We numerically integrated the equation of motion of dust particles using \editage{the} hydrodynamical simulation data. \editage{3D orbital calculations} were performed in the frame co-rotating with the planet (the local box in \Figref{fig:config}). Through each 3D orbital calculation, we sampled the positions and velocities of particles at \editage{specific} time intervals\editage{. We then} obtained the spatial distribution of dust influenced by the planet-induced gas flow in the local domain (Sect. \ref{sec:Spatial distribution of dust in the local domain}).}


\revsec{Although our orbital calculations were performed in the local domain, in reality, the dust particles distributed outside the domain as well, i.e., in the full azimuthal direction of a disk. \revtir{Thus, third, we assumed the uniform and Gaussian distributions of dust in the azimuthal and vertical directions outside the local domain of orbital calculations (Sect. \ref{sec:Spatial distribution of dust in the global domain}), and then calculated the azimuthally and vertically averaged drift velocities of dust (Sect. \ref{sec:Azimuthally and vertically average drift velocity of dust}) based on the obtained spatial distributions of dust both inside and outside the local domain of orbital calculations.}}

\rev{Finally, we compute\revsec{d} the dust surface density by incorporating the obtained azimuthally and vertically average\revsec{d} drift velocity into \editage{a} one-dimensional (1D) advection-diffusion equation (Sect. \ref{sec:Simulation-of-dust-surface-density}).}

\subsection{\revfor{Dimensionless unit system}}
Throughout all of our simulations, the length, times, velocities, and densities are normalized by the disk scale height, $H$,  reciprocal of the orbital frequency, $\Omega^{-1}$, sound speed $c_{\rm s}$, and  unperturbed gas density at the location of the planet, $\rho_{\rm disk}$, respectively. \editage{Section \ref{sec:Disk model} describes the relation between dimensionless and dimensional quantities for the steady accretion disk model.}

\revnin{While the mass of the disk gas is normalized by $\rho_{\rm disk}H^3$, our simulations which neglect the self-gravity of the gas allow us to introduce another normalization for the planetary mass \citep{Ormel:2015a} as given by}
 \begin{align}
     m=\frac{R_{\rm Bondi}}{H}=\frac{M_{\rm pl}}{M_{\rm th}},
 \end{align}
where $R_{\rm Bondi}=GM_{\rm pl}/c_{\rm s}^2$ is the Bondi radius of the planet, $M_{\rm pl}$ is the mass of the planet, $G$ is the gravitational constant, and $M_{\rm th}$ is the thermal mass of the planet \citep{goodman2001planetary}:
\begin{align}
    M_{\rm th}=\frac{c_{\rm s}^3}{G\Omega}=M_{\ast}\left(\frac{H}{r}\right)^3,
\end{align}
where $M_{\ast}$ is the mass of the central star \revfif{and $r$ is the orbital radius}. The Hill radius is given by $R_{\rm Hill}=(m/3)^{1/3}H$ in this unit. \AK{The relation between dimensionless planetary masses and dimensional ones for \revsec{a} \rev{steady viscous-accretion disk model is shown later in \Figref{fig:physical-quantities}a. \revsec{We focus on low-mass planets ($m\leq0.3$) which do not shape the global pressure gradient and carve a gas gap in a disk.\footnote{\revtir{In our dimensionless unit, \revfif{the pebble isolation mass is \revsix{given by $m_{\rm iso}\gtrsim0.6$} \citep{Lambrechts:2014,Bitsch:2018,ataiee2018much}}. When the Hill radius of the planet exceeds the disk scale height, a gas gap forms close to the planetary orbit. This happens when $m>3$ \citep{Lin:1993}.}}} The range of dimensionless planetary mass in this study is $m=0.03\text{--}0.3$, which corresponds to a planet with $\sim0.1\text{--}10\,M_{\oplus}$ orbiting a solar-mass star at $\sim1\text{--}10$ au.}}

\iffigure
\begin{figure*}
    \centering
    \includegraphics[width=\linewidth]{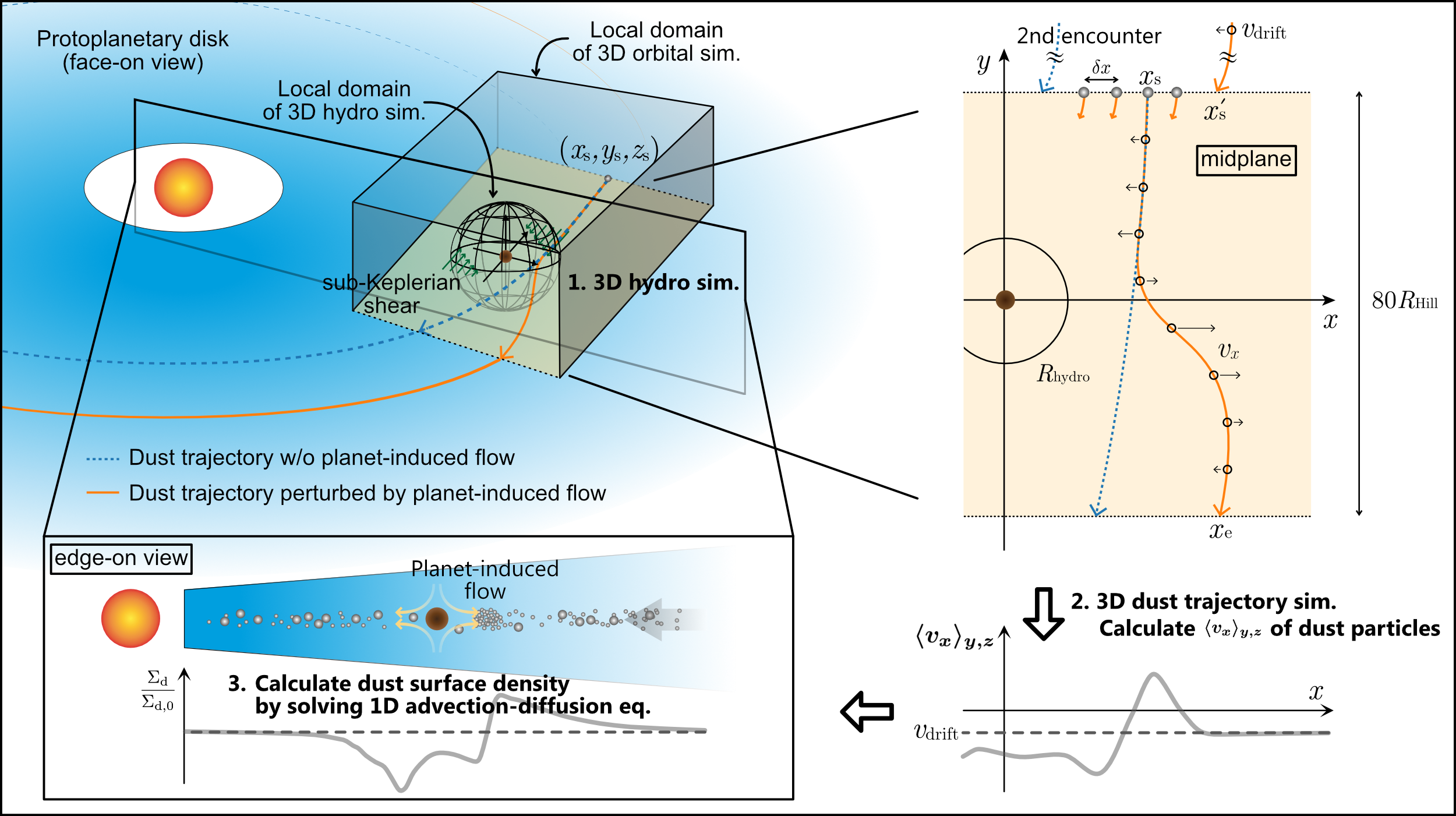}
    \caption{Schematic illustration of our model. We performed 3D hydrodynamical simulations on the spherical polar grid (the black sphere) hosting a planet located at its center (the brown-filled circle; \rev{Sect. \ref{sec:Three-dimensional hydrodynamical simulations (the local domain)}}). We consider the planet on a fixed circular orbit with the orbital radius, $a$. We adopt unperturbed sub-Keplerian shear flow at the outer boundary of the hydrodynamical simulations (the green arrows). We then calculated the trajectories of particles influenced by the planet-induced gas flow in the frame co-rotating with the planet (the local box; \rev{Sect. \ref{sec:Three-dimensional-orbital-calculation-of-dust-particles}). \rev{We used the limited part of the computational domain of hydrodynamical simulation, $R_{\rm hydro}$, in orbital calculations of dust particles (Sect. \ref{sec:Three-dimensional-orbital-calculation-of-dust-particles}).} The $x$- and $z$-coordinates of the starting point of particles, $x_{\rm s}$ and $z_{\rm s}$, are the parameters. The starting point of orbital calculation is beyond $R_{\rm hydro}$. \rev{The spatial interval of orbital calculation in the $x$-direction is $\delta x$ (Sect. \ref{sec:Spatial distribution of dust in the local domain}}). The $x$-coordinate of the particle at the edge of the domain of orbital calculation is $x_{\rm e}$. The $x$-coordinate of the particle after one synodical orbit, $x'_{\rm s}$, is given by \Equref{eq:drift distance} (Sect. \ref{sec:Existence of forbidden region}).} The schematic trajectories of particles are shown by the dashed blue (\rev{without the influence of planet-induced gas flow}) and \AK{solid} orange (perturbed by the planet-induced gas flow) lines. The open circles on the trajectory correspond to the position of a particle at \editage{specific} time intervals, $\Delta t(v_{y,0},z_{\rm s})$ \revfor{(\Equref{eq:delta t})}. \rev{The black arrows on the trajectories are} the velocities in the $x$-direction of a particle at each position, $v_{x}$. \editage{Note} that we do not follow trajectories outside the calculation domain of orbital calculations. \rev{Instead, we assumed the uniform and Gaussian distributions of dust particles in the azimuthal and vertical directions outside the local domain of orbital calculation, respectively, where the particles have the fixed velocity in the $x$-direction, $v_{\rm drift}$. Based on the obtained spatial distribution of dust in the global domain}, we calculate the azimuthally and vertically average\revsec{d} drift velocity, $\langle v_x\rangle_{y,z}$ (\Equref{eq:average-velocity}). Finally, we calculated the dust surface density by \rev{incorporating $\langle v_x\rangle_{y,z}$ into} the 1D advection-diffusion equation (\Equref{eq:advection-diffusion equation}).}
    \label{fig:config}
\end{figure*}
\fi

\subsection{\revsec{Three-dimensional hydrodynamical simulations (the local domain)}}\label{sec:Three-dimensional hydrodynamical simulations (the local domain)}
We performed nonisothermal, \revtir{inviscid} 3D hydrodynamical simulations of the gas of the protoplanetary disk around an embedded planet. \reveig{The inviscid assumption is justified in Sect. \ref{sec:Spatial distribution of dust in the local domain}.} \AK{The obtained gas flow data were used to simulate the orbits of dust particles \rev{influenced by the planet-induced gas flow (described later in Sect. \ref{sec:Numerical methods: Dust (local domain)})}.} 

\revsec{Our methods of hydrodynamical simulations are the same as described \editage{by} \cite{Kuwahara:2020a,Kuwahara:2020b}.} Our simulations were performed in a spherical polar coordinate co-rotating with a planet \editage{using the} Athena++ code\footnote{\revfif{https://github.com/PrincetonUniversity/athena}} \citep[\revfor{\Figref{fig:config};}][]{White:2016,stone2020athena++}. \revfor{Hereafter we denote $x$-, $y$- and $z$-direction as radially outwards (away from the star), azimuthal (the orbital direction of the planet), and vertical direction\editage{s}, respectively.} We \editage{used varying sizes} of the inner boundary, \rev{$R_{\rm inn}$, as a function of} the mass of the planet \AK{to follow \rev{a mass-radius relationship of} the non-gaseous part of the planet}:
\rev{
\begin{align}
R_{\rm inn}\simeq3\times10^{-3}m^{1/3}\,H\ \Biggl(\frac{\rho_{\rm pl}}{5\ {\rm g/}{\rm cm}^{3}}\Biggr)^{-1/3}\ \Biggl(\frac{M_{\ast}}{1\ M_{\odot}}\Biggr)^{1/3}\ \Biggl(\frac{r}{1\ {\rm au}}\Biggr)^{-1},\label{eq:planetaryradius}
\end{align}
where $\rho_{\rm pl}$ is the density of the embedded planet and $M_{\odot}$ is the solar mass.}\footnote{\revtir{We assumed $r=1$ au in \Equref{eq:planetaryradius}. The size of the inner boundary does not affect our results \citep{Kuwahara:2020a}. Therefore, when we convert the dimensionless planetary mass into \editage{a} dimensional one, we can use an arbitrary orbital radius regardless of a fixed $R_{\rm inn}$ (Sect. \ref{sec:Disk model}).}} A planet is embedded in an inviscid gas disk and is orbiting around a solar-mass star with \revsec{a} fixed orbital radius, $a$. The unperturbed gas velocity in the local frame is assumed to be the speed of the sub-Keplerian shear,
\begin{align}
    \bm{v}_{\rm g,\infty}=\left(-\frac{3}{2}x-\mathcal{M}_{\rm hw}\right)\bm{e}_{y},
\end{align}
where $\mathcal{M}_{\rm hw}$ is the Mach number of the headwind of the gas, $v_{\rm hw}$. \rev{The gas velocity has \editage{a} constant value, $\bm{v}_{{\rm g},\infty}$, at the outer edge of the computational domain of hydrodynamical simulation\revsec{s}. The headwind of the gas} is given by
\begin{align}
    v_{\rm hw}=\eta v_{\rm K},
\end{align}
where
\begin{align}
    \eta=-\frac{1}{2}\left(\frac{c_{\rm s}}{v_{\rm K}}\right)^2\frac{\mathrm{d}\ln \revnin{p}}{\mathrm{d}\ln r},\label{eq:eta}
\end{align}
is a dimensionless quantity characterizing the global pressure gradient of the disk gas. \rev{In \Equref{eq:eta},} $\revnin{p}$ is the pressure of the gas and $v_{\rm K}$ is the Keplerian velocity. \AK{The Mach number increases as the orbital distance increases for \rev{the steady viscous-accretion disk model (as shown later in \Figref{fig:physical-quantities}b). The range of the Mach number in this study is $\mathcal{M}_{\rm hw}=0.01\text{--}0.1$, which covers a wide range of the disk ($\lesssim100$ au).}} We \editage{have listed} our parameter sets in Table \ref{tab:1}.

\begin{table*}[htbp]
\caption{List of hydrodynamical simulations. The following columns give the simulation name, the size of the Bondi radius of the planet, the size of the Hill radius of the planet, the size of the inner boundary, the size of the outer boundary, the size of the limited part of the calculation domain (Sect. \ref{sec:Three-dimensional hydrodynamical simulations (the local domain)}), the Mach number of the headwind, \revten{the length of the calculation time,} and the regime of the planet-induced gas flow \citep[][see also Sect. \ref{sec:Three-dimensional planet-induced gas flow}]{Kuwahara:2020b}, respectively.}
\centering
\begin{tabular}{lcccccccc}\hline\hline
Name & $R_{\rm Bondi}$ & $R_{\rm Hill}$ & $R_{\rm inn}$ & $R_{\rm out}$ & $R_{\rm hydro}$ & $\mathcal{M}_{\rm hw}$ & \revten{$t_{\rm end}$} & flow regime\\ \hline
\texttt{m003-hw001,-hw003,-hw01} & 0.03 & 0.22 & 9.32$\times10^{-4}$ & 0.5 & 0.3 & 0.01, 0.03, 0.1 & \revten{50} & flow-shear, -shear, -headwind \\
\texttt{m01-hw001,-hw003,-hw01} & 0.1 & 0.32 & 1.39$\times10^{-2}$ & 5 & 3 & 0.01, 0.03, 0.1 & \revten{150} & flow-shear, -shear, -shear \\
\texttt{m03-hw001,-hw003,-hw01} & 0.3 & 0.46 & 2$\times10^{-2}$ & 5 & 5 & 0.01, 0.03, 0.1 & \revten{200} & flow-shear, -shear, -shear \\\hline
\end{tabular}
\label{tab:1}
\end{table*}

\subsection{\revsec{Dust \revtir{trajectory} simulations (the local domain)}}\label{sec:Numerical methods: Dust (local domain)}
\AK{To clarify the influence of the planet-induced gas flow on the radial drift of dust particles in a disk, \rev{we calculated dust trajectories in gas flow obtained by hydrodynamical simulations (Sect. \ref{sec:Three-dimensional-orbital-calculation-of-dust-particles}). We generated the spatial distribution of dust influenced by the planet-induced gas flow in the local domain (Sect. \ref{sec:Spatial distribution of dust in the local domain}).}}

\subsubsection{Three-dimensional orbital calculation of dust particles}\label{sec:Three-dimensional-orbital-calculation-of-dust-particles}

\AK{We calculated the trajectories of dust particles influenced by the planet-induced gas flow in the frame co-rotating with the planet (\Figref{fig:config}). \revfor{\editage{The $x$-, $y$- and $z$-directions are similar to those used in the hydrodynamical simulations.}} \editage{A majority} of our methods \editage{for} orbital calculations are the same as described in detail \editage{by} \cite{Kuwahara:2020a,Kuwahara:2020b}, where we studied \editage{the} accretion of dust particles onto a protoplanet embedded in a disk. We \editage{have described} the differences \editage{in the} numerical setting of orbital calculations with our previous works \revfif{later in Sect. \ref{sec:Numerical-settings:-3D-orbital-calculation}.}}

In our co-rotating frame, the $x$- and $y$-components of the initial velocity of dust particles are given by the drift equations \citep{Weidenschilling:1977a,Nakagawa:1986}:
\begin{align}
    v_{x,0}=&-\frac{2{\rm St}}{1+{\rm St}^2}\mathcal{M}_{\rm hw}\equiv v_{\rm drift},\label{eq:initial-vx}\\
    v_{y,0}=&-\frac{\mathcal{M}_{\rm hw}}{1+{\rm St}^2}-\frac{3}{2}x,\label{eq:initial-vy}
\end{align}
where ${\rm St}$ is the dimensionless stopping time of a dust particle, called the Stokes number,
\begin{align}
    {\rm St}=t_{\rm stop}\Omega.
\end{align}
We assume ${\rm St}=10^{-4}$--$10^0$. \AK{The relation between the Stokes numbers and dimensional particle sizes for \rev{\revsec{a} steady viscous-accretion disk model is shown later in \Figref{fig:physical-quantities}c.} \rev{The parameter range in this study is \revsec{${\rm St}\sim10^{-4}$--$10^{0}$}, which corresponds to the dust particles of \revsec{$\sim0.1$ mm--$1$ m} at $\sim1$--$10$ au for the steady\revsec{-}state viscous-accretion disk (\Figref{fig:physical-quantities}c).}} 

\rev{We numerically integrated the equation of motion of dust particles. \revfor{We used the gas velocity obtained from the hydrodynamical simulation to calculate the gas drag force acting on \editage{the} dust within the limited part of the computational domain, $R_{\rm hydro}$ \revfif{\citep[see][for details]{Kuwahara:2020a,Kuwahara:2020b}. The sizes of the limited part of the computational domains, $R_{\rm hydro}$, are listed in Table \ref{tab:1}.} We assumed the Stokes gas drag regime. \revfif{The choice of the gas drag regime \editage{did} not affect our results (discussed in Appendix \ref{sec:Gas drag regimes}).}} The $y$-coordinate of the starting point of particles is fixed at $|y_{\rm s}|=40R_{\rm Hill}$ \citep{Ida:1989}. The $x$- and $z$-coordinates of the starting point of particles, $x_{\rm s}$ and $z_{\rm s}$, are the parameters. The ranges of $x_{\rm s}$ and $z_{\rm s}$ are $x_{\rm s}\in[-5R_{\rm Hill},5R_{\rm Hill}]$ and $z_{\rm s}\in[0,1]$. \editage{Owing} to the symmetry of the system, we only consider $z\geq0$. We set the upper limit as $z=1$. At high altitudes, $z>1$, the dust density decreases significantly (described later in \Equref{eq:Gaussian distribution}). The ranges of $x_{\rm s}$ and $z_{\rm s}$ fully cover the size of the perturbed region, which can be scaled by the Bondi radius in our parameter sets ($m\in[0.03,0.3]$).}

\subsubsection{\rev{Spatial distribution of dust in the local domain}\label{sec:Spatial distribution of dust in the local domain}}
Through each orbital calculation, we \AK{sampled} the positions and velocities of the particles at \AK{fixed} small time intervals\editage{. We} then obtained the spatial distribution of particles in the local calculation domain \AK{as schematically illustrated in the upper right corner of \Figref{fig:config}}. 

\rev{We first describe the spatial intervals of \revfor{the starting points of the orbital calculations\editage{,} which determine the radial and vertical distributions of \editage{the} incoming dust} (\Figref{fig:config}).} We integrated the equation of motion for dust particles with their initial spatial intervals of $\delta x=0.01b$ in the $x$-direction and of $\delta z=0.01$ in the $z$-direction. \rev{\revfor{The maximum impact parameter of accreted particles in the unperturbed flow, $b$, is expressed by \Equref{eq:impact parameter} \revfif{in Appendix \ref{sec:Maximum impact parameter of accreted particle}.}}} \rev{The spatial interval in the $x$-direction,} $\delta x=0.01b$, has on the order of $\sim10^{-4}$--$10^{-3}\,[H]$.

\revsec{Next, we} \rev{describe the time interval for sampling the position and velocity of a particle to generate the dust spatial distribution \revfor{within a local domain of orbital calculation\revfif{s}}. The time interval is given by}
\rev{
\begin{align}
    \Delta t=\frac{L}{|v_{y,0}|N_{\rm loc}(z_{\rm s})},\label{eq:delta t}
\end{align}
}where $L=80R_{\rm Hill}$ (the size of the calculation domain) and \rev{$N_{\rm loc}(z_{\rm s})$ is the number of dust particles on an unperturbed orbit (without the influences of gravity of the planet and the planet-induced gas flow) in the local calculation domain (described later in \Equref{eq:Gaussian distribution}). The dependence of $\Delta t$ on $|v_{y,0}|$ means that the dust mass flux is constant in the $x$-direction of the local calculation domain. \editage{To} take into account the vertical distribution of incoming dust, we assumed the Gaussian distribution:
\begin{align}
    N_{\rm loc}(z_{\rm s})=N_{0}\exp\left[-\frac{1}{2}\left(\frac{z_s}{H_{\rm d}}\right)^2\right],\label{eq:Gaussian distribution}
\end{align}
where $N_{0}$ is the number of dust particles at the midplane.} We adopt $N_{0}=500$. \rev{In \Equref{eq:Gaussian distribution}, the scale height of dust particles, $H_{\rm d}$, is given by} \citep{Durbrulle:1995,Cuzzi:1993,Youdin:2007}:
\begin{align}
    H_{\rm d}=\left(1+\frac{\rm St}{\alpha_{\rm diff}}\frac{1+2{\rm St}}{1+{\rm St}}\right)^{-1/2}H,\label{eq:dust scale height}
\end{align}
where $\alpha_{\rm diff}$ is the dimensionless turbulent parameter, \rev{which describes turbulent diffusion in our model. \revfor{We assume $\alpha_{\rm diff}=10^{-5},\,10^{-4},\,\text{and}\,10^{-3}$.} \editage{Note} that the obtained spatial distribution of dust in the local domain deviates from the Gaussian distribution in the vertical direction because of the gas flow in the vertical direction.}

\revfif{We assumed \editage{the} inviscid fluid of an ideal gas and did not consider the effect of random motion of dust due to the turbulence in a series of hydrodynamical simulations and orbital calculations \citep{Kuwahara:2020a,Kuwahara:2020b}. The dimensionless turbulent parameter, $\alpha_{\rm diff}$, \editage{is} used when we consider the vertical distribution of dust (Sects. \ref{sec:Spatial distribution of dust in the local domain} and \ref{sec:Model for the global dust distribution}) and the turbulent diffusion of dust in the radial direction (described later in Sect. \ref{sec:Simulation-of-dust-surface-density}).} \reveig{As long as we focus on the range of the turbulence strength considered in this study, \revele{the turbulence likely affects the dust only.} \revten{The viscosity suppresses the 3D flow only when its $\alpha$-viscous coefficient is large, $\alpha_{\rm diff}\gtrsim10^{-2}$ \citep{Fung:2015,Lambrechts:2017}. We note that several previous studies \revele{reported another transition at even lower $\alpha_{\rm diff}$.} \cite{Lambrechts:2017} found the oscillational flow pattern at $\alpha_{\rm diff}\lesssim10^{-3}$. In our inviscid simulations, we also found the fluctuation \revele{after $10\,\Omega_{\rm K}^{-1}$, but its amplitude is small: <10\% in the $x$-component of the outflow velocity.} \cite{mcnally2019migrating} \revele{considered the planets with $m\gtrsim0.3$ and} found the growth of a set of Rossby wave instability vortices outside the planetary \revele{orbits} when $\alpha_{\rm diff}\lesssim10^{-4}$ in their \revele{2D} global hydrodynamical simulations. Though these vortices do not appear in our local hydrodynamical simulations, \revele{their influences on dust drift may need to be investigated in a future study}.}}

\subsection{\revtir{Model for  global dust distribution}}\label{sec:Model for the global dust distribution}
\rev{So far, we only consider\revsec{ed} the distribution of dust particles in the local domain of orbital calculation because we \revsec{did} not follow trajectories of dust outside the domain (Sect. \ref{sec:Three-dimensional-orbital-calculation-of-dust-particles}). In reality, the dust particles distribute outside the domain \revsec{as well}, i.e., in the full azimuthal direction of the disk. We then consider\revsec{ed} the spatial distribution of dust \revsec{outside the local domain of orbital calculation} (Sect. \ref{sec:Spatial distribution of dust in the global domain}). Based on the obtained spatial distribution of dust \revsec{both inside and outside the local domain of orbital calculation,} we calculate\revsec{d} the azimuthally and vertically average\revsec{d} drift velocity of dust (Sect. \ref{sec:Azimuthally and vertically average drift velocity of dust}).}

\subsubsection{\rev{Spatial distribution of dust in the global domain}}\label{sec:Spatial distribution of dust in the global domain}
We assumed the uniform distribution of dust particles \rev{in the azimuthal direction outside the \revsec{local domain of orbital calculation}, where the particles have the fixed velocity in the $x$-direction, $v_{\rm drift}$. We confirmed that the velocities of dust particles \editage{were} almost identical to that of the initial values at the edge of the local domain of orbital calculation (Eqs. (\ref{eq:initial-vx}) and (\ref{eq:initial-vy})). Thus, this assumption \editage{was} justified.}

To maintain consistency with orbital calculations, outside the domain, the number of dust particles on \rev{a hypothetical orbit (the orange solid line outside the \revsec{local domain of orbital calculation}, which is schematically illustrated in the upper left corner of \Figref{fig:config})} is \AK{given} by
\rev{
\begin{align}
    N_{\rm out}(z_{\rm s})&\simeq\Biggl(\frac{2\pi a}{L}-1\Biggr)\,N_{\rm loc}(z_{\rm s}),
\end{align}
where we assume $a\gg x$. As with the orbital calculations, we consider the ranges of $x\in[-5R_{\rm Hill},5R_{\rm Hill}]$ and $z\in[0,1]$. The spatial intervals of hypothetical orbits are $\delta x$ and $\delta z$ in the $x$- and $z$-directions, respectively. Thus, the "global box" of size $10R_{\rm Hill}\times(2\pi a-L)\times1\,[H^3]$ contains $\sum_{i=0}^{100}N_{\rm out}(i/\delta z)\times10R_{\rm Hill}/\delta x$ particles.} 



\subsubsection{\AK{\rev{Azimuthally and vertically} average\revsec{d} drift velocity of dust}}\label{sec:Azimuthally and vertically average drift velocity of dust}
\rev{Based on the obtained spatial distribution of dust \revsec{both inside and outside the local domain of orbital calculation} (Sect. \ref{sec:Spatial distribution of dust in the global domain}), we calculate\revsec{d} the azimuthally and vertically average\revsec{d} drift velocity of dust.} \rev{We divided the obtained spatial distribution of dust in the global domain into grids in the $x$-direction with  equal spacing, $\Delta x=10\times\delta x$.} \rev{The total number of sampling particles in each grid} \AK{is given by
\begin{align}
    N_{{\rm tot},i}=N_{{\rm loc},i}+N_{{\rm out},i},
\end{align}
where $N_{{\rm loc},i}$ is obtained from orbital calculations and $N_{{\rm out},i}$ is given by $N_{{\rm out},i}=N_{\rm out}\times(\Delta x/\delta x)$}. \rev{The subscript $i$ denotes the grid number. Each grid contains \rev{$N_{{\rm tot},i}\sim10^6$} \AK{sampling} particles.} We confirmed that the \AK{sample number} is \rev{large} enough to achieve numerical convergence.

 The azimuthally and vertically averaged $x$-component of the velocity on the $i$th grid is calculated by 
\AK{
\begin{align}
    \langle v_{x,i}\rangle_{y,z}&=\displaystyle\frac{\int_0^1\int_0^{2\pi} \rho_{\rm d}v_{x,i}\mathrm{d}y\mathrm{d}z}{\int_0^1\int_0^{2\pi} \rho_{\rm d}\mathrm{d}y\mathrm{d}z},\label{eq:raw average velocity}
\end{align}
where $\rho_{\rm d}$ is the density distribution of dust
\begin{align}
    \rho_{\rm d}=\rho_{{\rm d},0}\exp\left[-\frac{1}{2}\left(\frac{z_s}{H_{\rm d}}\right)^2\right],
\end{align}
with the midplane density, $\rho_{{\rm d},0}=1$. We numerically calculated \Equref{eq:raw average velocity} as
\begin{align}
    \langle v_{x,i}\rangle_{y,z}=\frac{N_{{\rm loc},i}}{N_{{\rm tot},i}}\frac{\sum_{j=1}^{N_{{\rm loc,}i}}v_{x,ij}}{N_{{\rm loc,}i}}+\frac{N_{{\rm out},i}}{N_{{\rm tot},i}}v_{\rm drift},\label{eq:average-velocity}
\end{align}
where the subscript $j$ denotes the number of particles \AK{in the $i$th grid. The \rev{azimuthally and vertically average\revsec{d}} drift velocity of dust is schematically illustrated in the \editage{lower-right} corner of \Figref{fig:config}.}
}



\iffigure
\begin{figure}
    \centering
    \includegraphics[width=\linewidth]{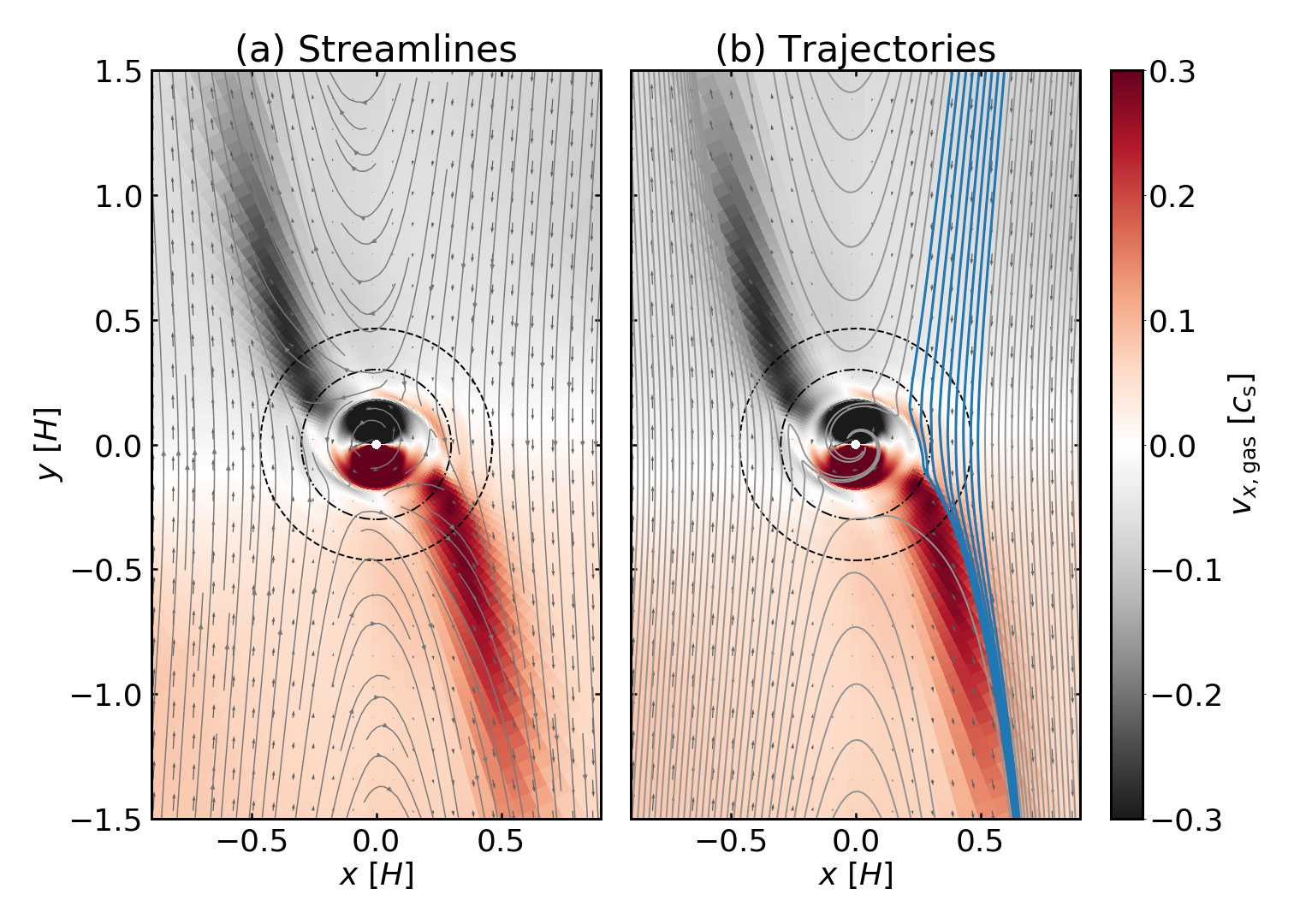}
    \caption{Streamlines \rev{of gas} (\textit{left}) and trajectories of dust particles with ${\rm St}=10^{-2}$ (\textit{right}) at the midplane of the disk. The flow structure around a planet obtained from \texttt{m03-hw001}. \AK{\rev{With this parameter set}, the planet-induced gas flow is in the flow-shear regime (see Sect. \ref{sec:Three-dimensional planet-induced gas flow}).} Color contour represents the flow speed in the $x$-direction, $v_{x,{\rm gas}}$. We set $z_{\rm s}=0$ for the orbital calculations. The dotted and dashed circles are the Bondi and Hill \AK{radii} of the planet, respectively. The thick blue lines in the right panel correspond to the trajectories of the dust particles which satisfies $x'_{\rm s}\geq x_{\rm s}$ (see \Figref{fig:config} and  Sect. \ref{sec:Existence of forbidden region} for details). \revfor{The 3D structure of the gas flow \editage{field} is shown in Appendix \ref{sec:Hydrodynamical regimes of the planet-induced gas flow}.}}
    \label{fig:streamlines-trajectories}
\end{figure}
\fi

\iffigure
\begin{figure}
    \centering
    \includegraphics[width=\linewidth]{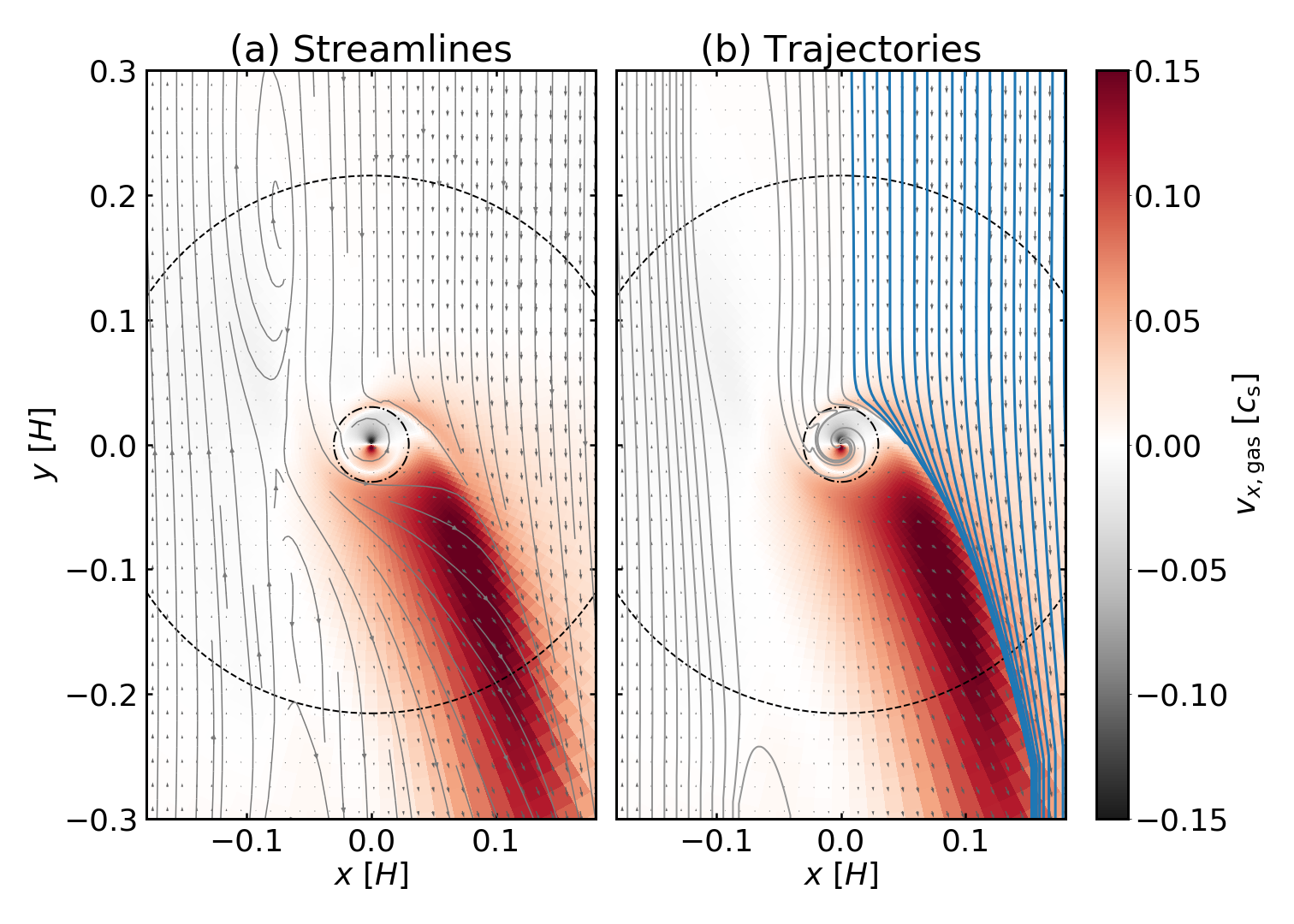}
    \caption{Same as \Figref{fig:streamlines-trajectories}, but the flow structure obtained from \texttt{m003-hw01}, and we set ${\rm St}=10^{-4}$ for the trajectories. \AK{In this figure, the planet-induced gas flow is in the flow-headwind regime (see Sect. \ref{sec:Three-dimensional planet-induced gas flow}).} \revfor{The 3D structure of the gas flow filed is shown in Appendix \ref{sec:Hydrodynamical regimes of the planet-induced gas flow}.}}
    \label{fig:streamlines_trajectories_m0.030_Mhw0.100}
\end{figure}
\fi

\iffigure
\begin{figure}
    \centering
    \includegraphics[width=\linewidth]{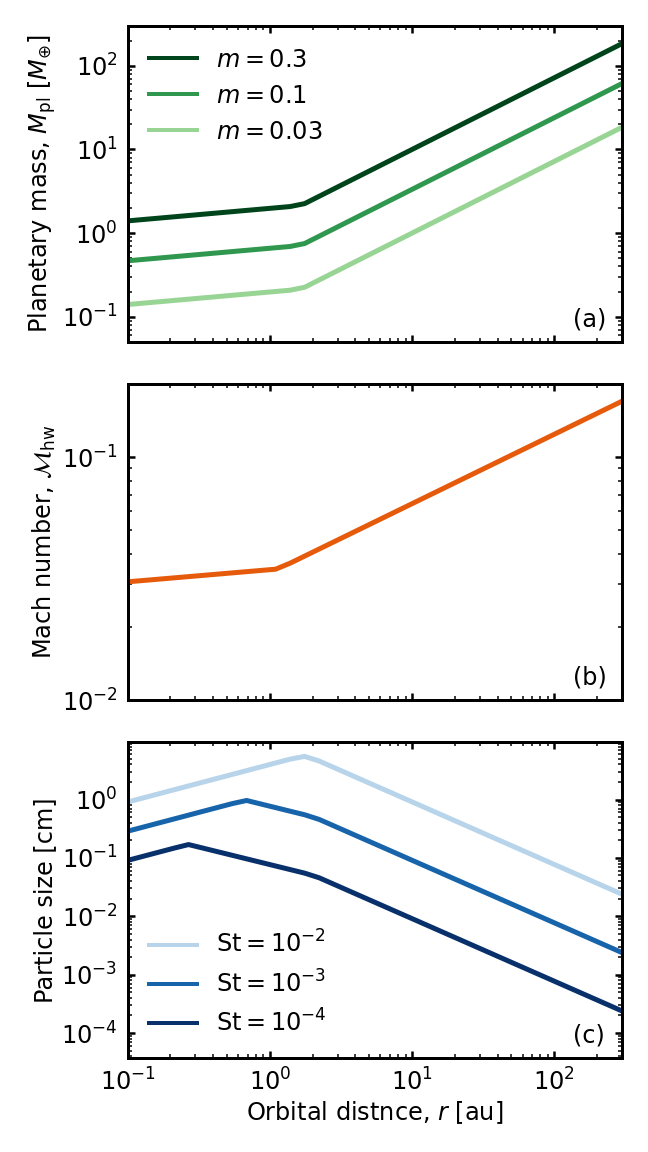}
    \caption{\AK{Relation} between the dimensionless quantities and dimensional ones. Each line represents the planetary mass (\textit{top}\revfor{; \Equref{eq:dimensional planetary mass}}), Mach number (\textit{middle}\revfor{; \Equref{eq:Mach number}}), and particle size (\textit{bottom}\revfor{; Eqs. (\ref{eq:Epstein-regime}) and (\ref{eq:Stokes-regime}})) as a function of the orbital radius. In \textit{panel c}, we only plotted ${\rm St}\leq10^{-2}$, which is the important parameter range in this study.}
    \label{fig:physical-quantities}
\end{figure}
\fi

\iffigure
\begin{figure*}
    \centering
    \includegraphics[width=\linewidth]{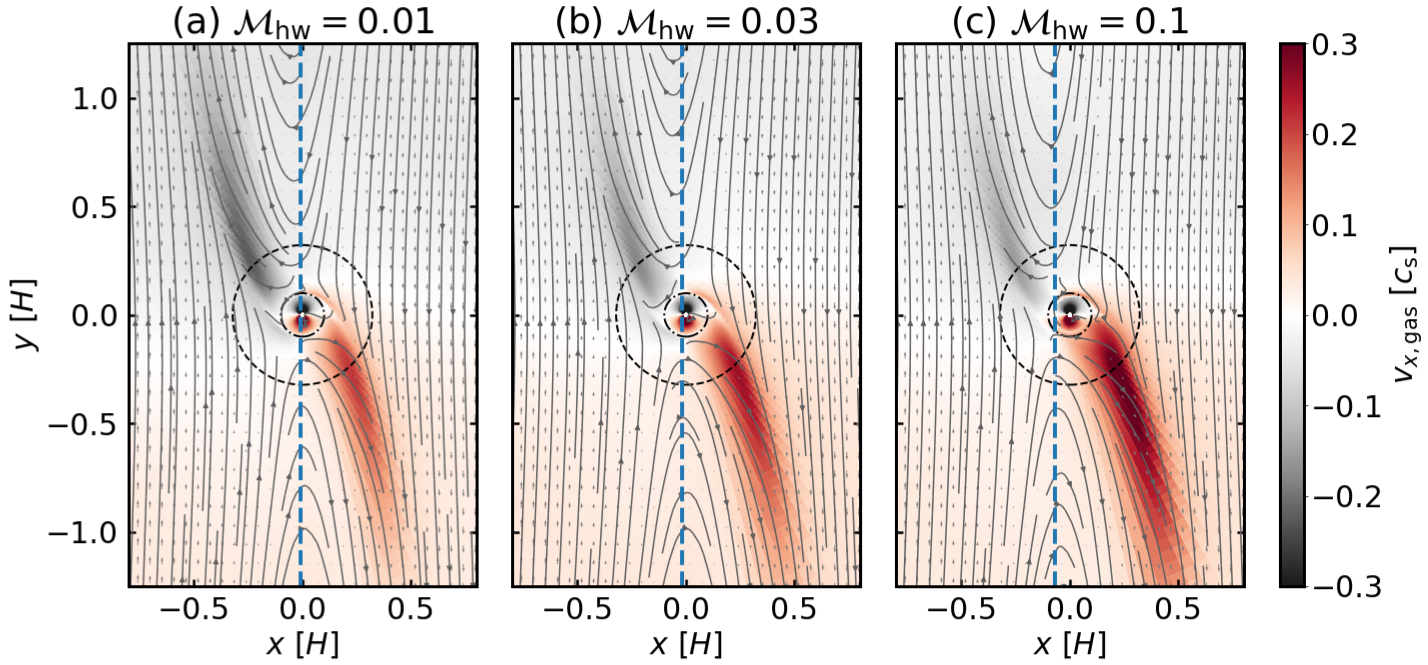}
    \caption{\revtir{Gas} flow structure around a planet. The results \editage{were} obtained from \texttt{m01-hw001} (\textit{left}), \texttt{m01-hw003} (\textit{middle}), and \texttt{m01-hw01} (\textit{right}). Color contour represents the flow speed in the $x$-direction, $v_{x,{\rm gas}}$. The solid lines correspond to the specific streamlines. The vertical dashed blue line corresponds to the corotation radius for the gas. The dotted and dashed circles are the Bondi and Hill radius of the planet, respectively.}
    \label{fig:streamlines_m0.1}
\end{figure*}
\fi
\subsection{\revfif{Simulation of dust surface density}}\label{sec:Simulation-of-dust-surface-density}
\rev{We computed the dust surface density by incorporating the azimuthally and vertically average\revsec{d} drift velocity into the 1D advection-diffusion equation \revfor{as schematically illustrated in the lower-left corner of \Figref{fig:config}. \editage{Instead of calculating the dust surface density directly from the obtained spatial distribution of dust, we solve the advection-diffusion equation for the following reason.}} The obtained spatial distribution of dust is a snapshot (Sect. \ref{sec:Spatial distribution of dust in the global domain}). \revsec{Though we considered the effect of turbulent diffusion in the vertical direction through \Equref{eq:dust scale height}, we did not consider it in the radial direction so far.}}

\AK{The dust surface density, $\Sigma_{\rm d}$, can be obtained by solving the following 1D advection-diffusion equation:
\begin{align}
    \frac{\partial \Sigma_{\rm d}}{\partial t}+\frac{\partial}{\partial x}\left(F_{\rm adv}+F_{\rm diff}\right)\label{eq:non-steady advection-diffusion equation}
    =0
\end{align}
, where $F_{\rm adv}= \langle v_x\rangle_{y,z}\Sigma_{\rm d}$ and $F_{\rm diff}=-\mathcal{D}\partial\Sigma_{\rm d}/\partial x$. $\langle v_x\rangle_{y,z}$ is given by \Equref{eq:average-velocity}\editage{,} and $\mathcal{D}$ is the diffusion coefficient for the dust \citep{Youdin:2007}:
\begin{align}
    \mathcal{D}=\frac{\alpha_{\rm diff}}{1+{\rm St}^2}.\label{eq:diffusion coefficient}
\end{align}
\revsec{\revnin{Because we focus on a radial range sufficiently narrow compared to the orbital radius of the planet, we assumed that the gas surface density is constant in these dust surface density simulations.} Thus, the gas surface density can be omitted in \Equref{eq:non-steady advection-diffusion equation}.} We assume\revsec{d} a steady\revsec{-}state and integrate\revsec{d} \revtir{\Equref{eq:non-steady advection-diffusion equation}} once, and then \editage{obtained}}
\begin{align}
    F_{\rm adv}+F_{\rm diff}=v_{\rm drift}\Sigma_{\rm d,0},\label{eq:advection-diffusion equation}
\end{align}
where $\Sigma_{\rm d,0}$ is the initial dust surface density.

The size of the calculation domain for solving \Equref{eq:advection-diffusion equation} is $x\in[-100R_{\rm Hill},100R_{\rm Hill}]$, \AK{which is much larger than that of the orbital calculations. Outside the domain of orbital calculations \revsec{in the radial direction}, $|x|>5R_{\rm Hill}$, we set $\langle v_x\rangle_{y,z}=v_{\rm drift}$ (see also \revfif{\ref{sec:Numerical settings: average drift velocity of dust}}).} \revtir{As a boundary condition, we assumed $\Sigma_{\rm d}=\Sigma_{\rm d,0}=1$.} 

\subsection{\revfif{Numerical settings}: dust trajectory simulations}\label{sec:Numerical-settings:-3D-orbital-calculation}
Most of our methods of orbital calculations are the same as described in detail \editage{by} \cite{Kuwahara:2020a,Kuwahara:2020b}, apart from the configuration of termination conditions. In \cite{Kuwahara:2020a,Kuwahara:2020b}, \AK{where we studied \editage{the} accretion of dust particles onto a protoplanet embedded in a disk}, we terminated orbital calculation \revtir{when either one of the following conditions was met: (1) the calculation time reached the sufficiently long time of computation, $10^4\,\Omega_{\rm K}^{-1}$, (2) a dust particle reached the edge of the calculation domain, $|y|\geq40R_{\rm Hill}$ (\Figref{fig:config}), or (3) accreted onto the planet, $R\leq2R_{\rm inn}$ \citep[see][for details]{Kuwahara:2020a,Kuwahara:2020b}.} However, we found that \AK{this treatment \editage{causes} several problems when \editage{studying} the influence of gas flow on dust drift} (Sect. \ref{sec:Simulation-of-dust-surface-density}).

First, we found that the $x$-component of the gas velocity, $v_{x,{\rm gas}}$, is \editage{significant in the vicinity of the planet}, $R\lesssim0.3R_{\rm Bondi}$ (Figs. \ref{fig:streamlines-trajectories} and \ref{fig:streamlines_trajectories_m0.030_Mhw0.100}). The particles that enter the Bondi region circulate close to the planet and then accrete onto the planet (Figs. \ref{fig:streamlines-trajectories}b and \ref{fig:streamlines_trajectories_m0.030_Mhw0.100}b). \AK{We found that the \rev{azimuthally and vertically} average\revsec{d} drift velocity of dust calculated by \Equref{eq:average-velocity} fluctuated significantly in the vicinity of the planetary orbit, \rev{$|x|\lesssim0.3R_{\rm Bondi}$,} due to the large $v_{x,{\rm gas}}$. To eliminate the fluctuation of the \rev{azimuthally and vertically} average\revsec{d} drift velocity, we assumed that the particles accreted onto the planet when $R\leq0.3R_{\rm Bondi}$ in all cases.}



Second, \AK{dust particles are trapped and circulate many times in the horseshoe region when the drift timescale of particles, $t_{\rm drift}\sim w_{\rm HS}/|v_{\rm drift}|$, where $w_{\rm HS}\simeq 2R_{\rm Bondi}$ is the width of the horseshoe region,\footnote{Our hydrodynamical simulations are performed in the local frame. The width of the horseshoe region obtained from our simulations differs from that obtained from global simulations \citep[e.g.,][]{Masset:2016}. See \cite{Kuwahara:2020b} for the discussion.} is comparable to the horseshoe libration timescale, $t_{\rm lib}\sim8\pi a/(3\Omega w_{\rm HS})$.}
\editage{We terminated orbital calculations to reduce the computational time} when the horseshoe turn \revtir{was} detected twice. 


\subsection{\revfif{Numerical settings: azimuthally and vertically averaged drift velocity of dust}}\label{sec:Numerical settings: average drift velocity of dust}
\rev{When we calculate\revsec{d} the azimuthally and vertically average\revsec{d} drift velocity of dust, $\langle v_x\rangle_{y,z}$, we found unexpected peaks in $\langle v_x\rangle_{y,z}$} at the positions away from the planetary orbit (outside the region influenced by the planet-induced outward gas flow, $|x|\gtrsim1$). These peaks \editage{are} caused by the density waves excited by an embedded planet. While our hydrodynamical simulations can resolve the local \revtir{gas} flow pattern in detail, they \revtir{do not handle} the global structure in the azimuthal direction in a disk. Since the density waves extend in the azimuthal direction, they are interrupted at the edge of the computational domain of our hydrodynamical simulations. \AK{We found that the location\revtir{s} of the unexpected peaks coincide with \revtir{those} where the density waves are interrupted.} \reveig{Appendix \ref{sec:Density waves excited by an embedded planet} shows the \revnin{locations of the outflows and the density waves.}}

\AK{Previous studies have reported that, in global disk simulations, the effects of density waves on dust drift \rev{were} minor} \citep{paardekooper2006dust,Muto:2009,rosotti2016minimum,dong2017multiple}. \revsec{Therefore}, we assumed that the dust particles have \editage{a} fixed velocity, $v_{\rm drift}$, outside the region influenced by the planet-induced outward gas flow. \editage{In this region,} the \revtir{artificial effect of the interrupted density waves} could be seen \AK{(schematically illustrated in the \editage{lower-right} corner of \Figref{fig:config})}. 


\subsection{\rev{Disk model}}\label{sec:Disk model}
\revtir{When we convert the dimensionless quantities into dimensional ones,} we consider the steady accretion disks with a constant turbulence strength \citep{Shakura:1973}, $\alpha_{\rm acc}$, including viscous heating due to the accretion of the gas and irradiation heating from the central star \citep[\Figref{fig:physical-quantities};][]{kusaka1970growth,nakamoto1994formation,oka2011evolution,Ida:2016}. \revfor{The dimensionless viscous alpha parameter, $\alpha_{\rm acc}$, describes the global efficiency of angular momentum transport of the gas \citep{Shakura:1973}. We assumed $\alpha_{\rm acc}=10^{-3}$.}

\revfor{The disk midplane temperature and the aspect ratio of the disk are given by Eqs. (\ref{eq:T vis})--(\ref{eq:aspect-ratio}) \revfif{in Appendix \ref{sec:Temperature and aspect ratio of the steady accretion disk}.}} \AK{In this disk model, the mass of the planet and the Mach number of the headwind can be described by (Figs. \ref{fig:physical-quantities}a and \ref{fig:physical-quantities}b):
\begin{align}
    M_{\rm pl}&=mM_{\ast}h^3\nonumber\\
    &\simeq\max\Biggl(6.6m\,\biggl(\frac{r}{1\,\text{au}}\biggr)^{3/20},\,4.6m\,\biggl(\frac{r}{1\,\text{au}}\biggr)^{6/7}\Biggr)\,M_{\oplus},\label{eq:dimensional planetary mass}
\end{align}
\begin{align}
    \mathcal{M}_{\rm hw}=\frac{\eta v_{\rm K}}{c_{\rm s}}&=-\frac{h}{2}\frac{\mathrm{d}\ln \revnin{p}}{\mathrm{d}\ln r}\nonumber\\
    &\simeq\max\Biggl(0.0344\,\biggl(\frac{r}{1\,\text{au}}\biggr)^{1/20},\,0.0333\,\biggl(\frac{r}{1\,\text{au}}\biggr)^{2/7}\Biggr),\label{eq:Mach number}
\end{align}
where we assume $\mathrm{d}\ln \revnin{p}/\mathrm{d}\ln r\simeq-2.55$ for the viscous region and $\mathrm{d}\ln \revnin{p}/\mathrm{d}\ln r\simeq-2.78$ for the irradiation region \citep{Ida:2016}.
}

In a steady accretion disk, the gas surface density is given by
\begin{align}
    \Sigma_{\rm g}=\frac{\dot{M}_{\ast}}{3\pi\alpha_{\rm acc}\reveig{r^2}h^2\Omega}.\label{eq:gas surface density}
\end{align}

The stopping time of the dust particle depends on the thermal structure of the disk. Thus, the Stokes number is given by \citep[\Figref{fig:physical-quantities}c;][]{Ida:2016}:
\begin{empheq}[left={{\rm St}=\empheqlbrace}]{alignat=2}
&\displaystyle\frac{\sqrt{2\pi}\rho_{\bullet}s}{\Sigma_{\rm g}}, \quad &(\text{Epstein regime}: s\leq\frac{9}{4}\lambda)\label{eq:Epstein-regime}\\
&\displaystyle\frac{4\rho_{\bullet}\sigma_{\rm mol} s^{2}}{9\mu m_{\rm H}h},  \quad &(\text{Stokes regime}: s\geq\frac{9}{4}\lambda),\label{eq:Stokes-regime}
\end{empheq}
where $\rho_{\bullet}$ is the internal density of the particle, $s$ is the radius of the particle, $\lambda$ is the mean free path of the gas,  $\lambda=\mu m_{\rm H}/\rho_{\rm g}\sigma_{\rm mol}$ with $\mu,\,m_{\rm H}$ and $\sigma_{\rm mol}$ being the mean molecular weight, $\mu=2.34$, the mass of the proton, and the molecular collision cross-section, $\sigma_{\rm mol}=2\times10^{-15}\,\text{cm}^2$ \citep{Chapman:1970,Weidenschilling:1977a,Nakagawa:1986}. 

\section{Results}\label{sec:result}
\iffigure
\begin{figure*}
    \centering
    \includegraphics[width=\linewidth]{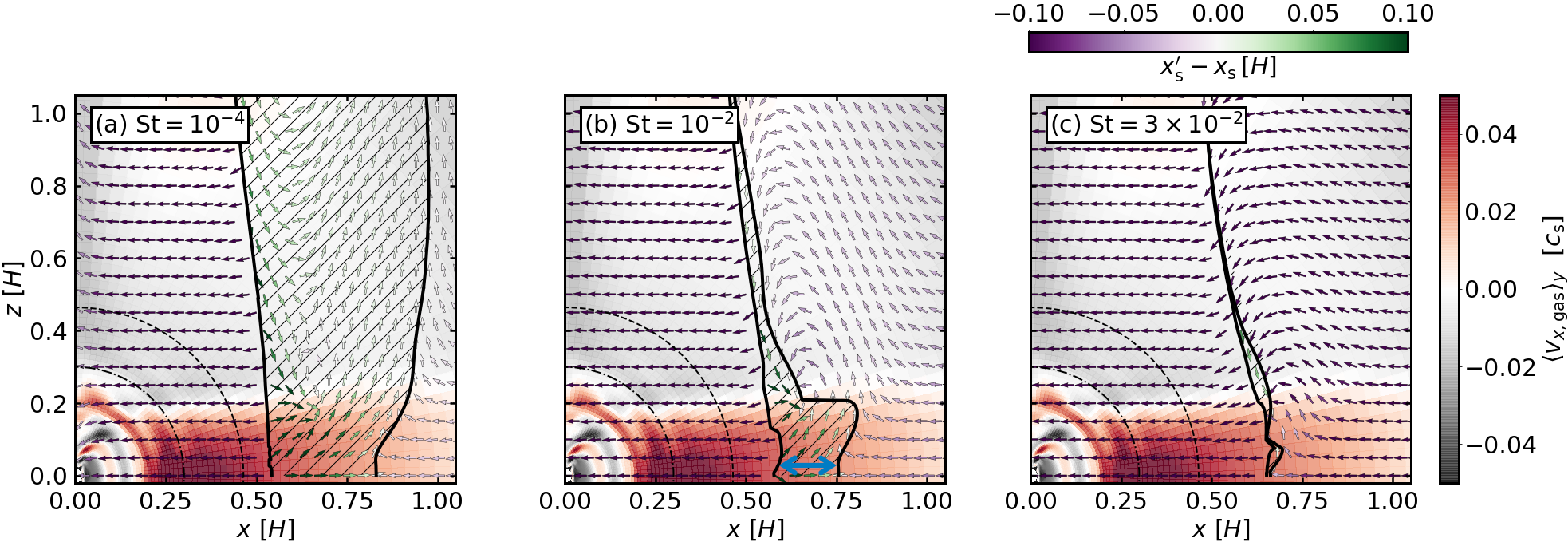}
    \caption{\revtir{\revfif{Vector field of dust velocities} in the cases with ${\rm St}=10^{-4}$ (\textit{left panel}), $10^{-2}$ (\textit{middle panel}), and $3\times10^{-2}$ (\textit{right panel}) under the common gas flow obtained from \texttt{m03-hw001}.} Color contour from red to black represents the \AK{$x$-component of the \revtir{gas} flow velocity averaged in the $y$-direction} within the calculation domain of hydrodynamical simulation around a planet, $\langle v_{x,{\rm gas}}\rangle_{y}$. Arrows represent the direction of movement of dust particles. Color contour \revtir{of the arrows} from green to purple represents the distance \editage{of drift of} a dust particle after one synodical orbit. \rev{The \revfif{hatched} region enclosed by black curves corresponds to the forbidden region}. \revfor{The two-headed blue arrow in \textit{panel b} can be compared to the region shown by the thick blue lines in \Figref{fig:streamlines-trajectories}b.} The dotted and dashed lines are the Bondi and Hill \AK{radii} of the planet, respectively. \AK{Thanks to the symmetry of the system,} we only focus on the region where $z\geq0$. \revfor{We discuss the dust motion in the $x$-$z$ plane in detail in Appendix \ref{sec:Dust motion within the forbidden region}.}}
    \label{fig:forbidden_region_m0.300_Mhw0.010}
\end{figure*}
\fi

\iffigure
\begin{figure}
    \centering
    \includegraphics[width=\linewidth]{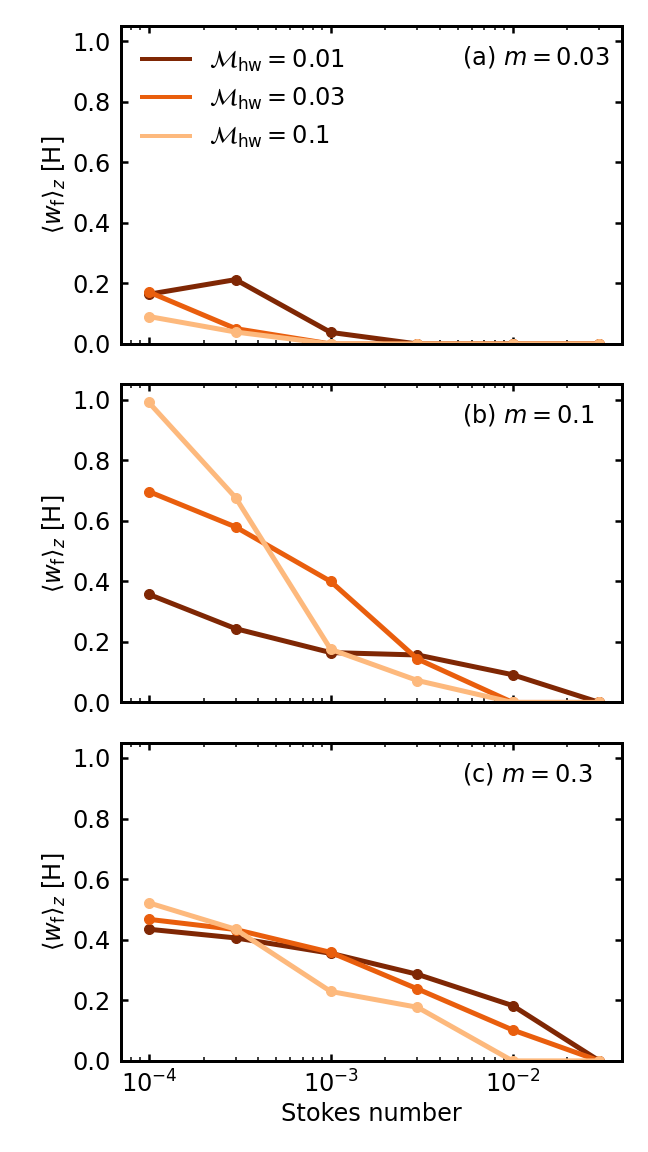}
    \caption{\rev{Vertically} average\revsec{d} width of the forbidden region, $\langle w_{\rm f}\rangle_{z}$. Different colors correspond to different $\mathcal{M}_{\rm hw}$. We adopt $\alpha_{\rm diff}=10^{-4}$.}
    \label{fig:<w>}
\end{figure}
\fi

\iffigure
\begin{figure}
    \centering
    \includegraphics[width=\linewidth]{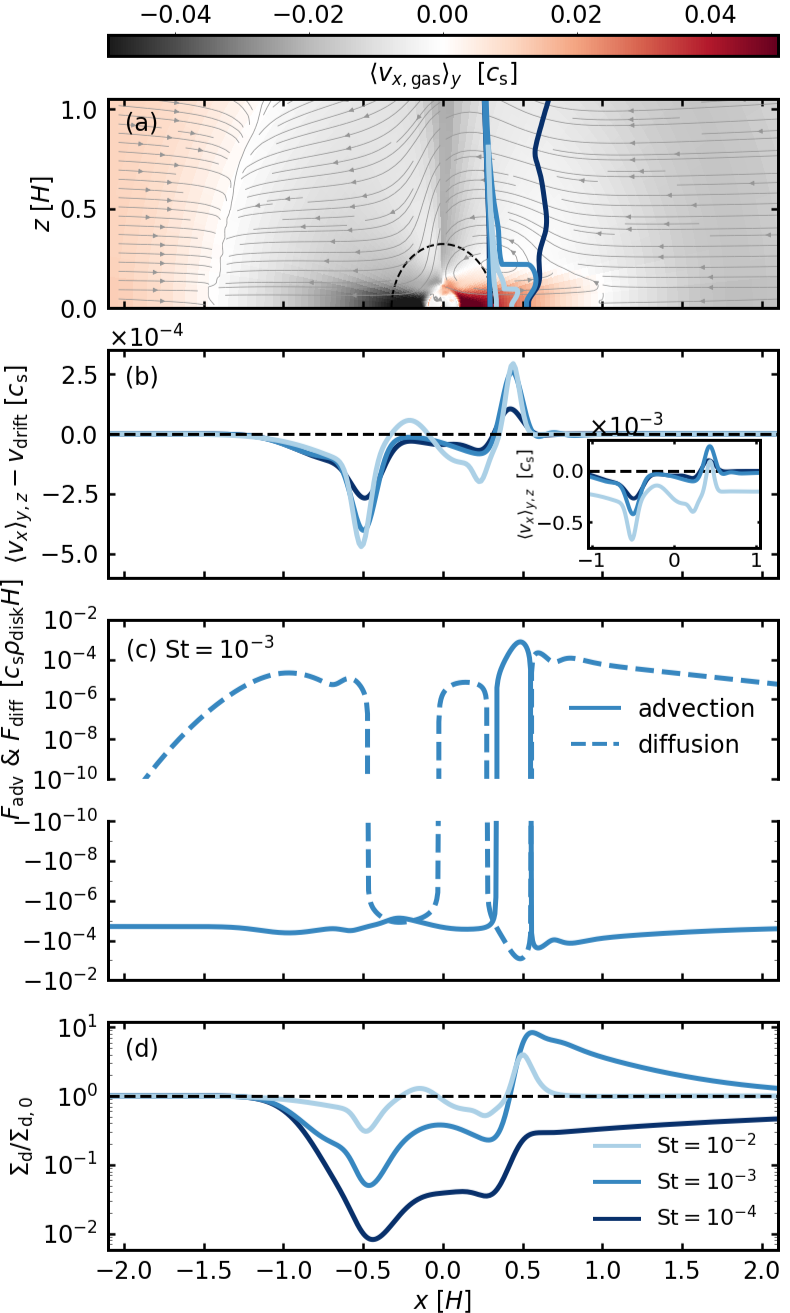} 
    \caption{\AK{\revtir{Gas} flow structure averaged in the $y$-direction within the calculation domain of hydrodynamical simulation} around a planet (\textit{panel a}), \rev{the difference of the azimuthally and vertically average\revsec{d} velocity of dust particles from the unperturbed steady\revsec{-}state drift velocity, $\langle v_x\rangle_{y,z}-v_{\rm drift}$} (\textit{panel b}), \AK{advective (solid lines) and diffusive (dashed lines) flux of dust with ${\rm St}=10^{-3}$ in the steady\revsec{-}state (\textit{panel c})}, and the dust surface density (\textit{panel d}). \rev{\textit{Panels a, b, and d:} different colors correspond to different Stokes numbers.} \textit{Panel a}: result obtained from \texttt{m01-hw001}. The gray solid lines correspond to the streamlines. The dashed circle is the Hill radius. \rev{The regions enclosed by blue curves} correspond to the forbidden regions. \AK{The horizontal dashed lines in \textit{panels b and d} correspond to \rev{$\langle v_{x}\rangle_{y,z}=v_{\rm drift}$} and $\Sigma_{\rm d}/\Sigma_{{\rm d},0}=1$.} \rev{The figure displayed at the \editage{lower-right} of \textit{panel b} shows $\langle v_x\rangle_{y,z}$.} \textit{Panels b--d}: we adopt $\alpha_{\rm diff}=10^{-5}$. \AK{To improve legibility, we only plot the region where $|x|\leq2$.}}
    \label{fig:dust_ring_and_gap}
\end{figure}
\fi

\iffigure
\begin{figure}
    \centering
    \includegraphics[width=\linewidth]{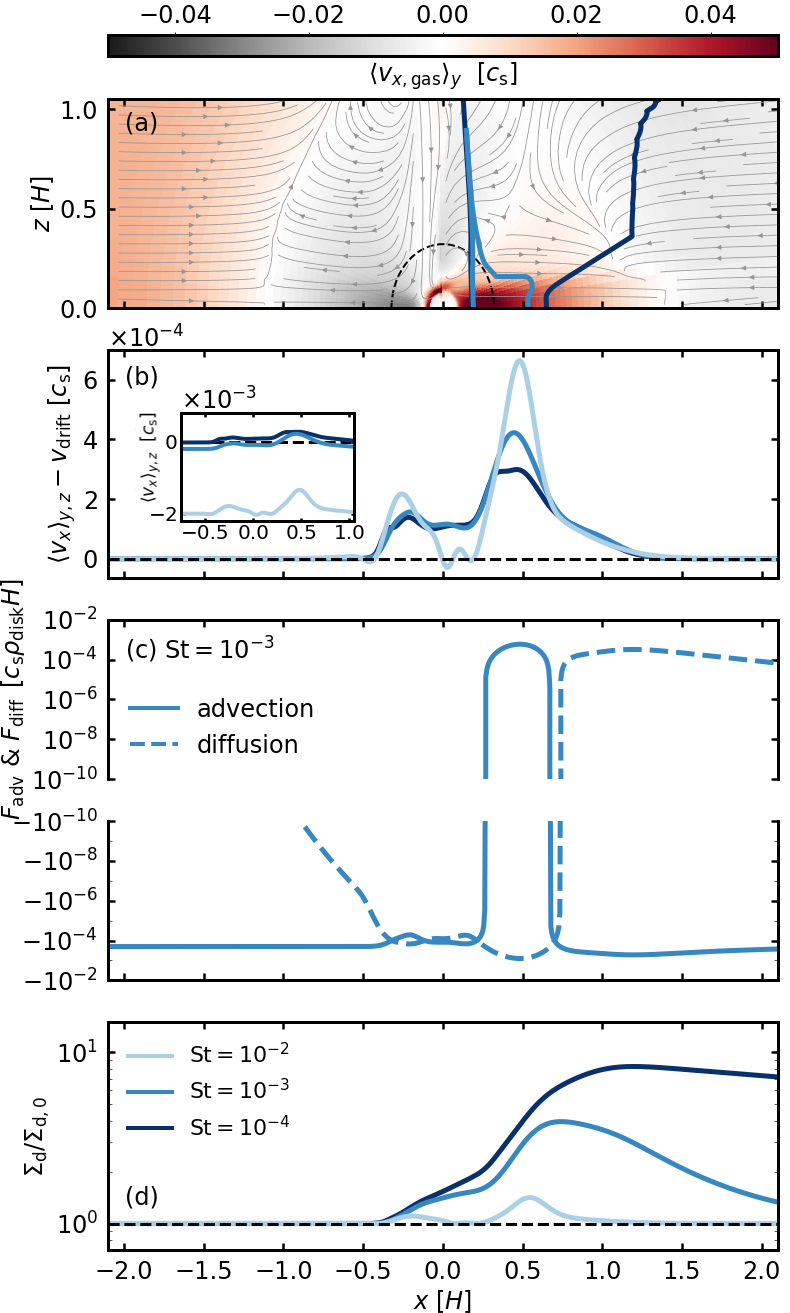}
    \caption{Same as \Figref{fig:dust_ring_and_gap}, but \editage{the} result obtained from \texttt{m01-hw01} in \textit{panel a}, and we adopt $\alpha_{\rm diff}=10^{-4}$ in \textit{panels b--d}. Since we only plot the region where $|x|\leq2$, the outer part of the ring is interrupted.}
    \label{fig:dust_ring_only}
\end{figure}
\fi

\iffigure
\begin{figure*}
    \centering
    \includegraphics[width=\linewidth]{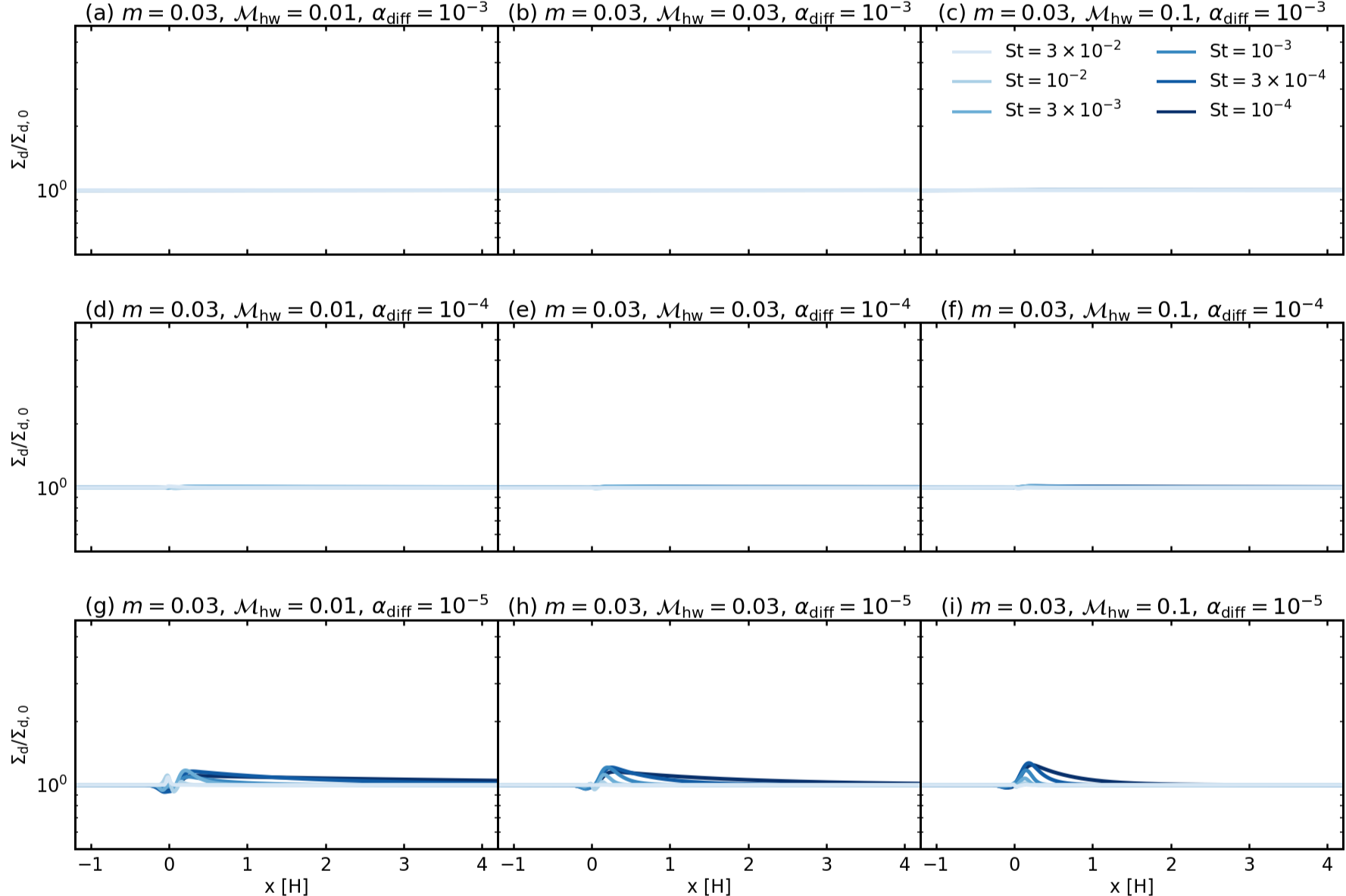}
    \caption{Dust surface density with $m=0.03$ for the different Stokes numbers, the Mach numbers, and the turbulent parameters. Different colors correspond to different Stokes numbers. \revsec{The minimum and the maximum values of the dust surface density are summarized in Figs. \ref{fig:Contour_min} and \ref{fig:Contour_max}}.}
    \label{fig:Sigma_sum_m0.030}
\end{figure*}
\fi

\iffigure
\begin{figure*}
    \centering
    \includegraphics[width=\linewidth]{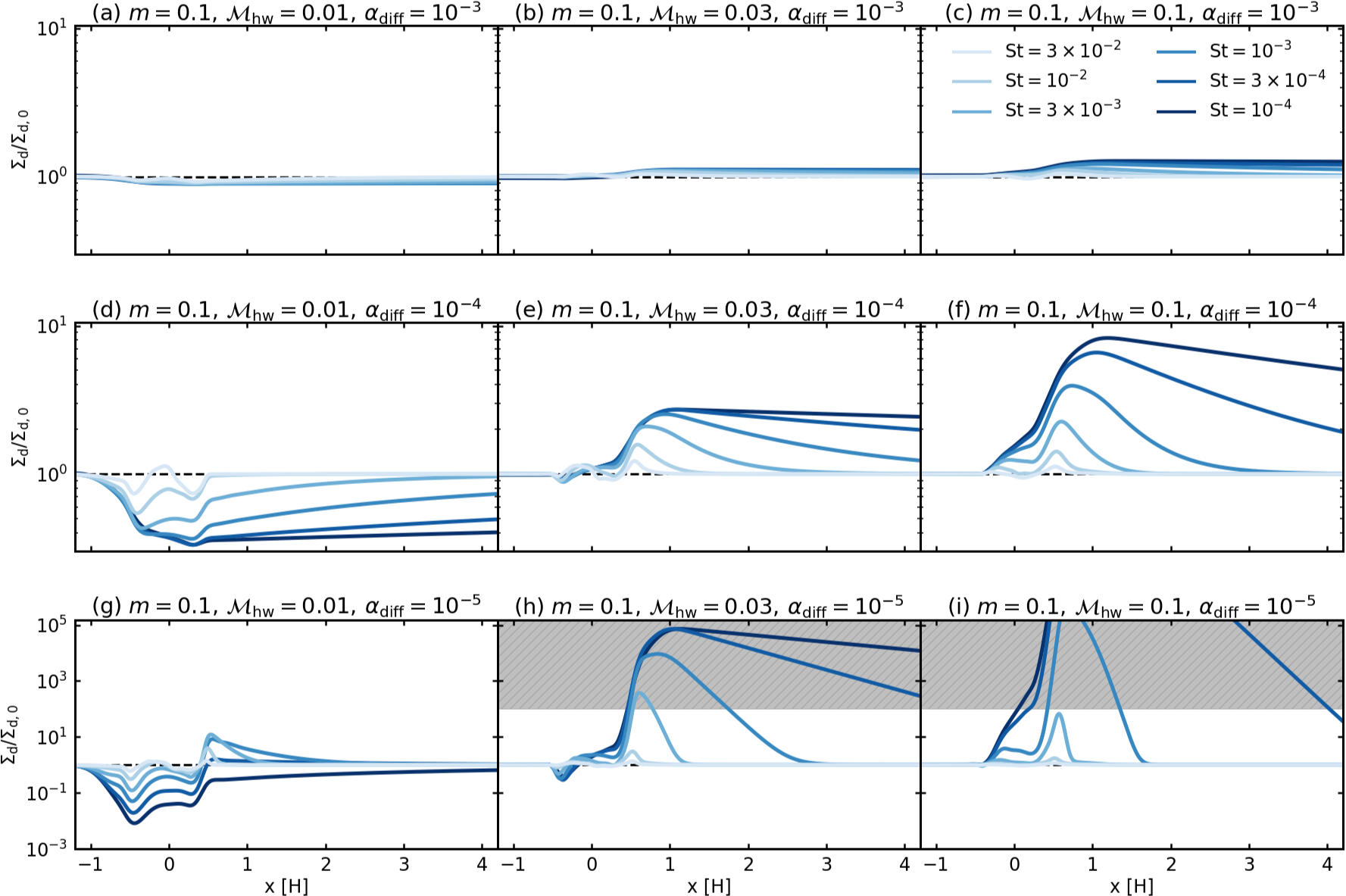}
    \caption{Same as \Figref{fig:Sigma_sum_m0.030}, but with $m=0.1$. The gray-shaded region represents an excessive accumulation of dust particles, where the dust-to-gas ratio exceeds the unity, assuming the initial value of $\Sigma_{\rm d,0}/\reveig{\Sigma_{\rm g}}=0.01$. Since we only plot the region where $-1\leq x\leq4$, the outer part of the ring is interrupted. \rev{The dashed black line corresponds to \revsec{$\Sigma_{\rm d}/\Sigma_{{\rm d},0}=1$.}}}
    \label{fig:Sigma_sum_m0.100}
\end{figure*}
\fi

\iffigure
\begin{figure*}
    \centering
    \includegraphics[width=\linewidth]{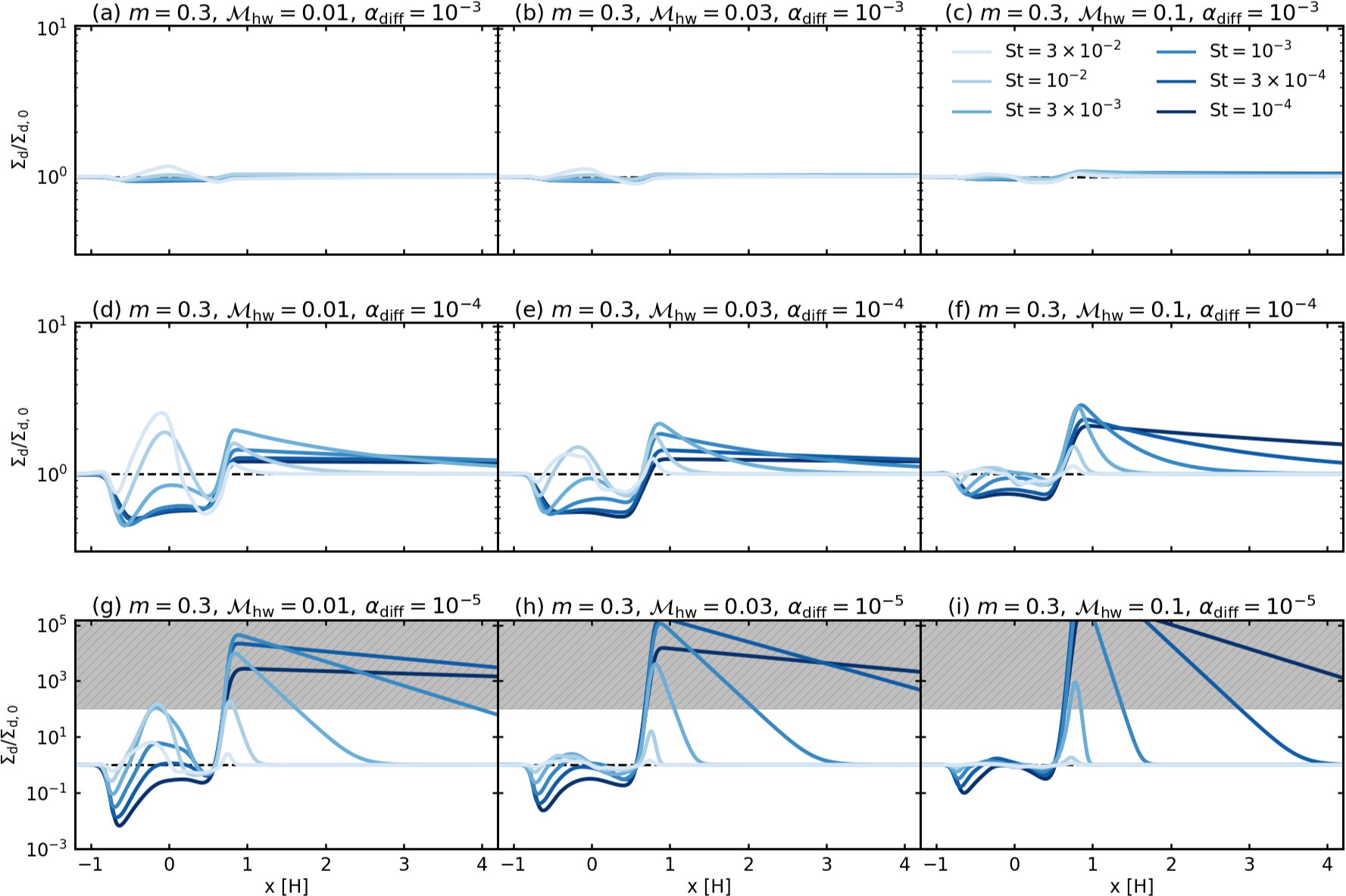}
    \caption{Same as \Figref{fig:Sigma_sum_m0.030}, but with $m=0.3$. The gray-shaded region represents an excessive accumulation of dust particles, where the dust-to-gas ratio exceeds the unity, assuming the initial value of $\Sigma_{\rm d,0}/\reveig{\Sigma_{\rm g}}=0.01$. Since we only plot the region where $-1\leq x\leq4$, the outer part of the ring is interrupted. \rev{The dashed black line corresponds to the initial value.}}
    \label{fig:Sigma_sum_m0.300}
\end{figure*}
\fi

\iffigure
\begin{figure*}
    \centering
    \includegraphics[width=\linewidth]{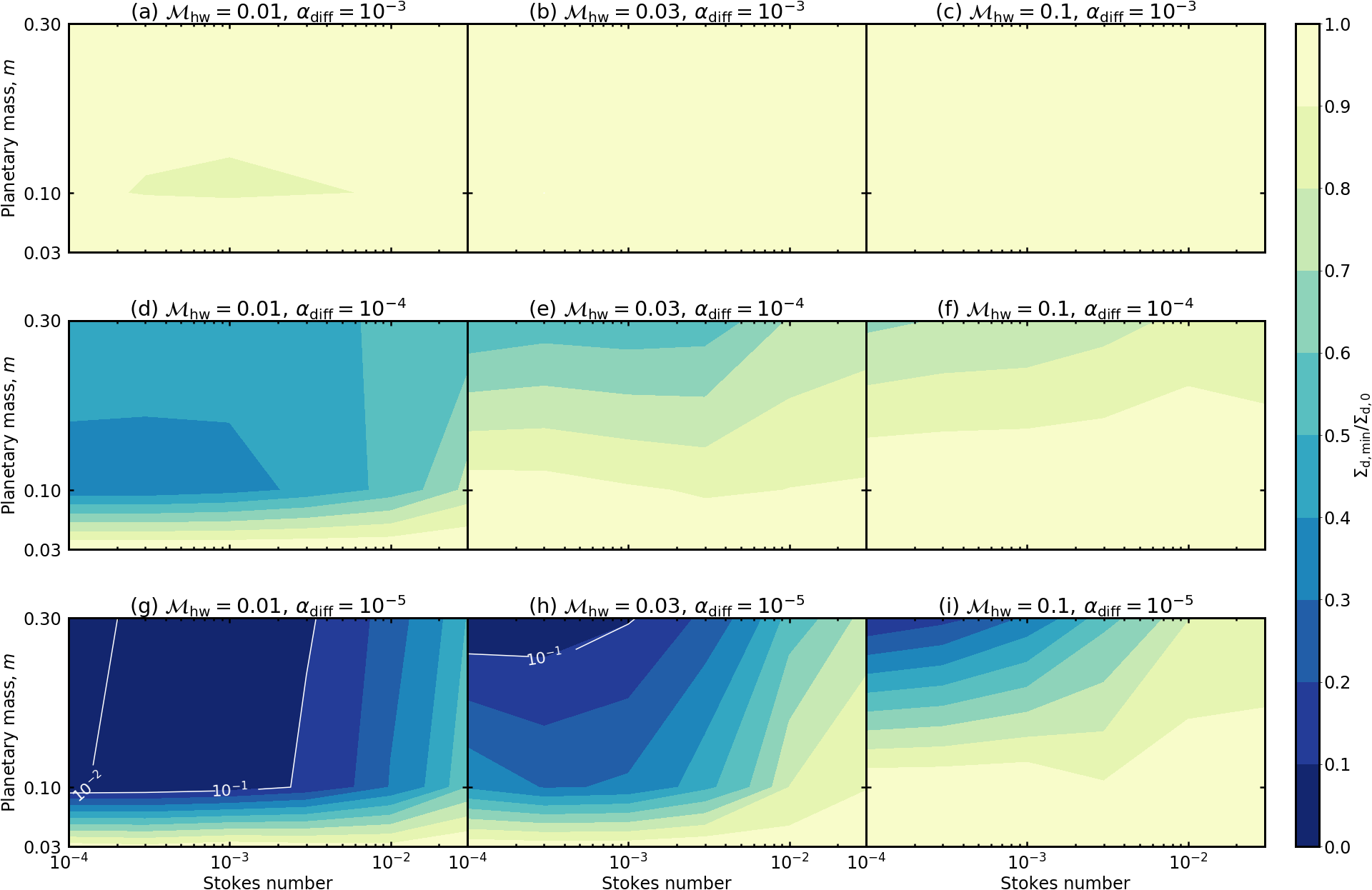}
    \caption{\revsec{Minimum dust surface density, $\Sigma_{\rm d,min}$ as a function of the planetary mass and the Stokes number. Different \revtir{rows} correspond to different turbulent parameters, $\alpha_{\rm diff}$. Different \revtir{columns} correspond to different Mach numbers, $\mathcal{M}_{\rm hw}$.}}
    \label{fig:Contour_min}
\end{figure*}
\fi

\iffigure
\begin{figure*}
    \centering
    \includegraphics[width=\linewidth]{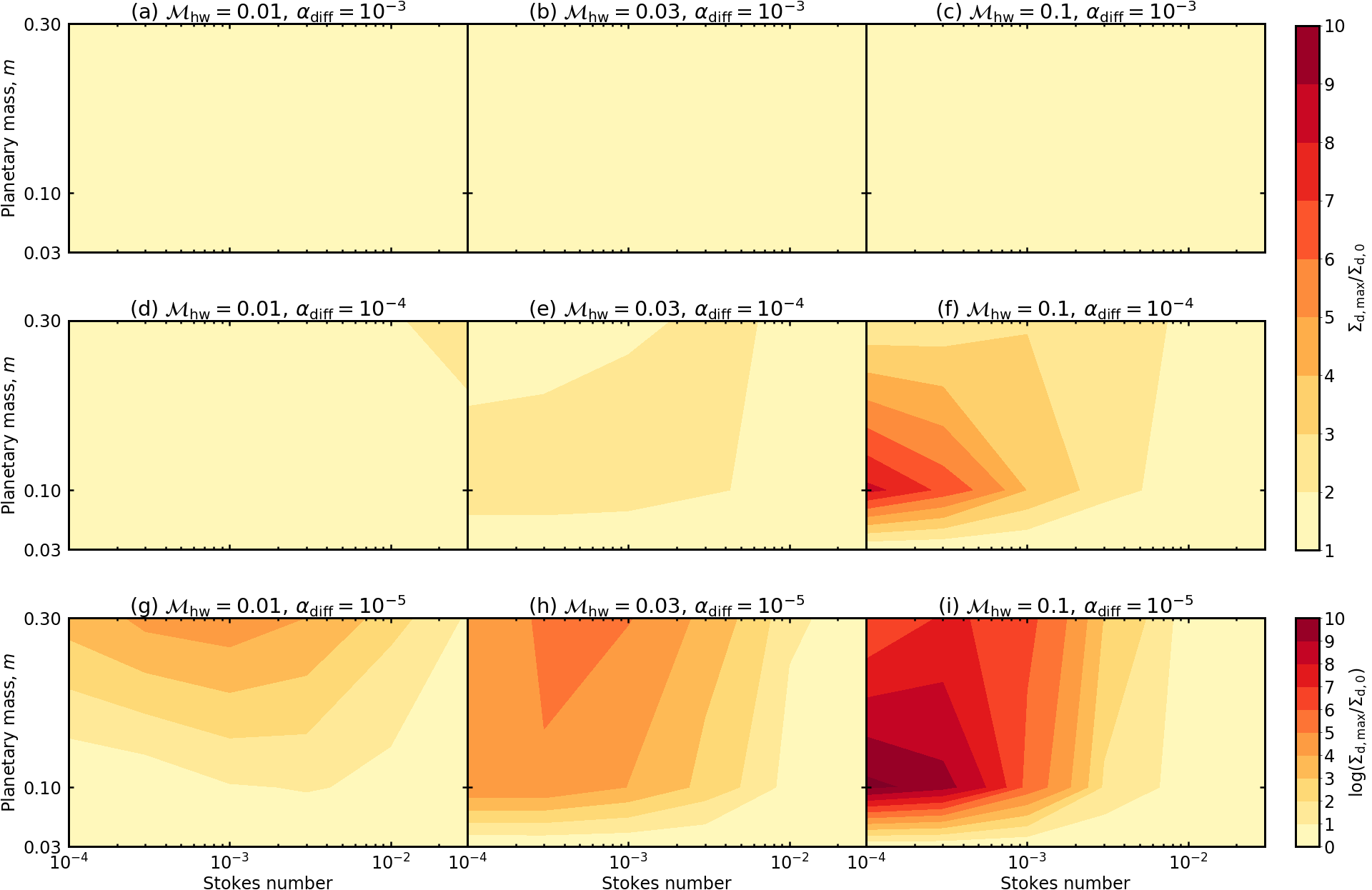}
    \caption{\revsec{Same as \Figref{fig:Contour_min}, but panels show the maximum dust surface density, $\Sigma_{\rm d,max}$, as a function of the planetary mass and the Stokes number. We note that the color contours are on a linear scale in \textit{panels a--f}, but on a log scale in \textit{panels g--i}.}}
    \label{fig:Contour_max}
\end{figure*}
\fi

\subsection{Results overview}
The main subject of this study is to clarify the influence of the planet-induced gas flow on the \AK{radial} distribution of dust particles in a disk. \editage{Section \ref{sec:Three-dimensional planet-induced gas flow} summarizes} the characteristic 3D structure of the planet-induced gas flow obtained by 3D hydrodynamical simulations. \revsec{In Sects. \ref{sec:Existence of forbidden region} and \ref{sec:Width of forbidden region}, we show the results of our orbital calculations. \editage{Section \ref{sec:Influence of planet-induced gas flow on the dust surface density} presents} the results obtained by solving \editage{the} 1D advection-diffusion equation for dust surface density.} 

\subsection{Three-dimensional planet-induced gas flow}\label{sec:Three-dimensional planet-induced gas flow}
The planet-induced gas flow has a complex 3D structure. \AK{A substantial amount of gas from the disk enters the Bondi \revsec{and} Hill sphere\revsec{s} at high latitudes (inflow), and exits through the midplane region of the disk (outflow).} The 3D flow structure depends on $\mathcal{M}_{\rm hw}$ \citep{Ormel:2015b,Kurokawa:2018}. In our previous work, we classified the planet-induced gas flow into the flow-shear and the flow-headwind regimes \citep[Figs. \ref{fig:streamlines-trajectories} and \ref{fig:streamlines_trajectories_m0.030_Mhw0.100};][]{Kuwahara:2020b}. \revfor{These flow regimes are important \editage{for understanding} the tendency of simulation results of dust surface density (Sect. \ref{sec:Influence of planet-induced gas flow on the dust surface density}).}

\revfif{Qualitatively, the planet-induced gas flow has a vertically rotational symmetric structure in the flow-shear regime\editage{. The} outflow occurs in the second and fourth quadrants of the $x$-$y$ plane (\Figref{fig:streamlines-trajectories}a). The symmetry is broken in the flow-headwind regime, where the outflow occurs only in the fourth quadrant of the $x$-$y$ plane (\Figref{fig:streamlines_trajectories_m0.030_Mhw0.100}a). In both regimes the planet-induced gas flow affects the motion of dust particles \citep[][see also Figs. \ref{fig:streamlines-trajectories}b and \ref{fig:streamlines_trajectories_m0.030_Mhw0.100}b]{Kuwahara:2020a,Kuwahara:2020b}. The 3D structure of the planet-induced gas flow is shown in Appendix \ref{sec:Hydrodynamical regimes of the planet-induced gas flow}.}

\AK{The transition from the flow-shear to the flow-headwind regime occurs when the dimensionless planetary mass \editage{drops} below the flow transition mass \citep{Kuwahara:2020b}:
\begin{align}
    m_{\rm t,flow}\simeq\frac{4}{15}\mathcal{M}_{\rm hw}.\label{eq:flow transition mass}
\end{align}
When $m\leq m_{\rm t,flow}$, the planet-induced \revtir{gas} flow is in the flow-headwind regime.} In our parameter set, only the case of \texttt{m003-hw01} is classified as the flow-headwind regime, otherwise \editage{classified} as the flow-shear regime based on the definition in \cite{Kuwahara:2020b} (\AK{see also} \Tabref{tab:1}). The flow transition mass was originally derived to distinguish these hydrodynamical regimes. 

The transition of the \revtir{gas} flow field from the flow-shear to the flow-headwind regime occurs smoothly. We found that the flow transition mass is not applicable for the classification of \rev{the effects on dust drift\editage{---which is the focus of this study---}but it is still useful to understand the tendency of our results. In this study,} the \revtir{gas} flow regime should be considered only as a qualitative guide to whether it is "shear-dominated" (two outflows occur both inside and outside the planetary orbit) or "headwind-dominated" (outflow occurs dominantly outside the planetary orbit). 

\revfif{Figure \ref{fig:streamlines_m0.1} shows the dependence of the gas flow structure around the planet with $m=0.1$ on $\mathcal{M}_{\rm hw}$. In this figure, the planet-induced gas flow is in the flow-shear regime based on the definition in \cite{Kuwahara:2020b}. As shown in \Figref{fig:streamlines_m0.1}, the outflow speed toward \editage{the inside (outside) of} the planetary orbit becomes weak (strong) as $\mathcal{M}_{\rm hw}$ increases.}
\subsection{\rev{Existence} of \revsec{the} forbidden region}\label{sec:Existence of forbidden region}
\AK{We first \revfif{show} that \revfor{the gas flow induced by low-mass planets affect the radial drift of \editage{the} dust particles}.} 
Given a dust particle launched at $(x_{\rm s},y_{\rm s})$, the $x$-coordinate of a particle after one synodical orbit at $y=y_{\rm s}$ can be described by
\begin{align}
    x'_{\rm s}\simeq x_{\rm e}-\Bigl(2\pi a-L\Bigr)\times\frac{v_{\rm drift}}{v_{y,0}},\label{eq:drift distance}
\end{align}
where $x_{\rm e}$ is the $x$-coordinate of a particle at the edge of the orbital calculation domain \AK{(upper right corner of \Figref{fig:config}). Equation (\ref{eq:drift distance}) \editage{is valid} because the velocity of dust is identical to that of the unperturbed value at the edge of the domain of orbital calculation (Sect. \ref{sec:Spatial distribution of dust in the local domain}).}  We then set the following simple criteria: (1) when $x'_{\rm s}<x_{\rm s}$, the dust particles continue the radial drift. (2) When $x'_{\rm s}\geq x_{\rm s}$, the \editage{planet-induced gas flow inhibits the radial drift. We only use the starting and ending points of an orbital calculation at this stage.}

\rev{We found that a forbidden region exists outside the planetary orbit, where the radial drift of dust particles is inhibited (\revfor{\Figref{fig:forbidden_region_m0.300_Mhw0.010}}).} Figure \ref{fig:forbidden_region_m0.300_Mhw0.010} shows \revfif{the vector field of dust velocities influenced by the planet-induced gas flow} in the $x$-$z$ plane. \rev{The mass of the planet and  Mach number of the headwind are $m=0.3$ and $\mathcal{M}_{\rm hw}=0.01$, respectively. We considered the Stokes numbers of ${\rm St}=10^{-4},\,10^{-2},\,\text{and}\,3\times10^{-2}$.} In \revtir{these cases}, the planet-induced gas flow is in the flow-shear regime. \rev{A significant outward drift of dust particles can be seen near the midplane (dark green arrows in \Figref{fig:forbidden_region_m0.300_Mhw0.010})\editage{,} because the outflow has the maximum speed near the midplane region of the disk.} \revfif{We \editage{have discussed} the motion of dust within the forbidden region in detail in Appendix \ref{sec:Dust motion within the forbidden region}.}


\AK{Similar to the flow-shear regime, \rev{we confirmed that} a forbidden region \rev{exists} outside the planetary orbit in the flow-headwind regime \rev{(see Sect. \ref{sec:Width of forbidden region})}.}

\subsection{Width of \revsec{the} forbidden region}\label{sec:Width of forbidden region}
There is a clear dependence of the \AK{width of the} forbidden region on the Stokes number. \AK{As shown in \Figref{fig:forbidden_region_m0.300_Mhw0.010}, the width of the forbidden region is wider for small Stokes number.}

Figure \ref{fig:<w>} shows the changes of \AK{the \rev{vertically} average\revsec{d} width of the forbidden region weighted by the dust number density}, $\langle w_{\rm f} \rangle_{z}$, as a function of the Stokes number for different planetary masses and Mach numbers. \rev{The vertically average\revsec{d} width of the forbidden region was} calculated by
\begin{align}
    \displaystyle\langle w_{\rm f} \rangle_{z}=\frac{\int_{0}^1 \rho_{\rm d} w_{\rm f}(z)\,\mathrm{d}z}{\int_{0}^1 \rho_{\rm d}\mathrm{d}z},
\end{align}
where $w_{\rm f}(z)$ is the width of the forbidden region as a function of $z$. \AK{Same as \Equref{eq:raw average velocity},} we only consider $0\leq z\AK{[H]}\leq1$. \AK{Overall trends indicate that} $\langle w_{\rm f} \rangle_{z}$ is a decreasing function of the Stokes number\editage{,} because the influences of gas flow are weaker for the large Stokes number. \reveig{We note that the forbidden region does not exist in the limit of ${\rm St}\rightarrow0$. \revnin{The inflow to the Bondi sphere of the planet balances the outflow from it.} This replenishment occurs through the meridional circulation of the gas \citep[e.g.,][]{Fung:2015}. When ${\rm St}\rightarrow0$, dust particles cycle along with the replenishment of the gas \revnin{and, consequently with the outflow}, meaning that the forbidden region does not exist in this limit. On the other hand, if the upper layer of the disk is dust-poor, we expect that the forbidden region does exist. This occurs when the dust scale height is smaller than the disk scale height, $H_{\rm d}<H$. We discuss again the required conditions in our model later in Sect. \ref{sec:Dependence on turbulence strength}.}

For the larger Stokes numbers, ${\rm St}>3\times10^{-2}$, the forbidden region does not exist in our parameter space\rev{s} (\AK{$m\in[0.03,0.3]$ and $\mathcal{M}_{\rm hw}\in[0.01,0.1]$}). \editage{The following sections} \revfor{show} only the results for ${\rm St}\leq3\times10^{-2}$. 

\subsection{Influence of planet-induced gas flow on the dust surface density}\label{sec:Influence of planet-induced gas flow on the dust surface density}

\revfif{The essential points of the previous sections are summarized in \Figref{fig:config}.} So far we \editage{have} discussed the influence of the planet-induced gas flow on the radial drift of dust particles without the turbulent diffusion \revsec{in the radial direction (Sects. \ref{sec:Existence of forbidden region} and \ref{sec:Width of forbidden region}, \AK{where we only used the starting and ending point\rev{s} of an orbital calculation}).} \editage{This section shows the steady-state dust surface density obtained by solving the} 1D advection-diffusion equation (\Equref{eq:advection-diffusion equation}). We found that the dust surface density profiles are different in the flow-shear \AK{(shear-dominated)} and flow-headwind (\AK{headwind-dominated}) regimes.\footnote{Again, the following results cannot be directly classified based on the \editage{hydrodynamical} regimes introduced by \cite{Kuwahara:2020b} because \AK{the transition of the \revtir{gas} flow field from the flow-shear to the flow-headwind regime occurs smoothly}. We use the \revtir{gas} flow regimes as a qualitative guide. The important point is whether the dominant outflow occurs inside or outside the planetary orbit.} Hereafter, we refer to \AK{the regions where dust is depleted and accumulated as "dust gap" and "dust ring," respectively.} 


\subsubsection{Dust \AK{ring and gap} formation by planet-induced gas flow: \AK{flow-shear regime}}\label{sec:Dust ring-gap formation by planet-induced gas flow: flow-shear regime}

When the outflow occurs dominantly both outside and inside the planetary orbit, \AK{that is, \rev{when} the planet-induced gas flow is in the flow-shear regime,} the dust is accumulated outside the planetary orbit (dust ring) and depleted around the planetary orbit (dust gap; \Figref{fig:dust_ring_and_gap}d). \revsec{Panels a--d of \Figref{fig:dust_ring_and_gap} show} the \revtir{gas} flow structure \AK{averaged in the $y$-direction within the calculation domain of hydrodynamical simulation (\texttt{m01-hw001}; \revsec{\Figref{fig:dust_ring_and_gap}a})}, \rev{the difference of the azimuthally and vertically average\revsec{d} drift velocity of dust particles from the unperturbed steady\revsec{-}state drift velocity} (\Equref{eq:average-velocity}; \revsec{\Figref{fig:dust_ring_and_gap}}b), \AK{advective and diffusive flux of dust} (\revsec{\Figref{fig:dust_ring_and_gap}c}), and the steady\revsec{-}state dust surface density (\revsec{\Figref{fig:dust_ring_and_gap}d}), respectively. \AK{We set $\alpha_{\rm diff}=10^{-5}$ in Figs. \ref{fig:dust_ring_and_gap}b-d.} We \editage{overplotted} the forbidden regions for different Stokes numbers in \Figref{fig:dust_ring_and_gap}a.

\AK{We found that the \rev{azimuthally and vertically average\revsec{d}} drift velocity of dust particles has positive and negative peaks outside and inside the planetary orbit (\Figref{fig:dust_ring_and_gap}b). The positive peak is located at the position overlapping the forbidden region. The negative peak is caused by the outflow toward \editage{the inside of} the planetary orbit, where the \rev{inward} drift of dust particles is enhanced.}

\editage{The advective flux dominates in the distant region far from the planet, $|x|\gg1$} (\Figref{fig:dust_ring_and_gap}c). The dust particles drift inward to the central star with the steady\revsec{-}state velocity, $v_{\rm drift}$, and thus the dust surface density has the fixed value, $\Sigma_{\rm d}=\Sigma_{{\rm d},0}$.

\AK{The dust particles accumulate at the location where $\langle v_x\rangle_{y,z}$ switches from negative to positive due to the outflow toward \editage{the outside of} the planetary orbit, which is shown by the solid lines of ${\rm St}=10^{-2}$ and $10^{-3}$ in \Figref{fig:dust_ring_and_gap}d. The gradient of the dust surface density caused by dust accumulation results in an outward (inward) diffusive flux on the right (left) side across the peak of the dust surface density (\Figref{fig:dust_ring_and_gap}c). The tail of the dust ring extends beyond the width of the forbidden region.}

\AK{The dust particles pass through the forbidden region by turbulence diffusion and are transported inside the planetary orbit (\Figref{fig:dust_ring_and_gap}c). Since the outflow occurs inside the planetary orbit in the flow-shear regime, the inward advective flux is enhanced by the outflow. As a result, the dust surface density decreases around the planetary orbit, \editage{followed by the formation of the dust gap.}
}

\AK{We found that only \editage{the} dust gap formation occurs in \editage{a few} cases (e.g., ${\rm St}=10^{-4}$ in \Figref{fig:dust_ring_and_gap}d)\editage{, such as when} the amplitude of the positive peak in the \rev{azimuthally and vertically average\revsec{d} drift velocity of dust} is smaller than that of the negative peak.}

\AK{The width and depth of the dust gap are larger for \rev{smaller Stokes numbers}. The typical width of the dust gap is $\sim1$ $[H]$ in most cases, which corresponds to the width of the region influenced by the planet-induced outflow (e.g., \Figref{fig:streamlines-trajectories}). In \Figref{fig:dust_ring_and_gap}, the width of the dust ring has on the order of $\sim0.1\text{--}1\,[H]$, but it strongly depends on the Stokes number and the turbulence strength (shown later in Figs. \ref{fig:Sigma_sum_m0.100} and \ref{fig:Sigma_sum_m0.300}, see Sect. \ref{sec:Dependence on turbulence strength}).}

\subsubsection{Dust ring formation by planet-induced gas flow: \AK{flow-headwind regime}}\label{sec:Dust ring formation by planet-induced gas flow: flow-headwind regime}
\AK{When the outflow occurs dominantly outside the planetary orbit (flow-headwind regime), the dust is accumulated outside the planetary orbit without the depletion of dust around the planetary orbit\editage{,} as shown in \Figref{fig:dust_ring_only}d. Similar to \Figref{fig:dust_ring_and_gap}, \revsec{panels a--d of \Figref{fig:dust_ring_only} show} the \revtir{gas} flow structure averaged in the $y$-direction within the calculation domain of hydrodynamical simulation (\texttt{m01-hw01}; \revsec{\Figref{fig:dust_ring_only}a}), the \rev{difference \editage{between} the azimuthally and vertically average\revsec{d} drift velocity of dust particles and the unperturbed steady\revsec{-}state drift velocity} (\revsec{\Figref{fig:dust_ring_only}}b), advective and diffusive flux of dust (\revsec{\Figref{fig:dust_ring_only}}c), and the dust surface density (\revsec{\Figref{fig:dust_ring_only}}d), respectively. We set $\alpha_{\rm diff}=10^{-4}$ in Figs. \ref{fig:dust_ring_only}b-d.}

\AK{In the flow-headwind regime, the \rev{azimuthally and vertically average\revsec{d}} drift velocity of dust particles has only a positive peak, which leads to dust accumulation outside the planetary orbit (Figs. \ref{fig:dust_ring_only}b and d). \rev{\editage{Note} that $\langle v_x\rangle_{y,z}$ for ${\rm St}=10^{-2}$ in \Figref{fig:dust_ring_only} is always negative (see the small panel displayed in \Figref{fig:dust_ring_only}b).} Even when the drift speed of dust is always negative, the dust particles accumulate slightly outside the planetary orbit (${\rm St}=10^{-2}$; \Figref{fig:dust_ring_only}d). This is caused by the traffic jam of dust that results from changes in $\langle v_x\rangle_{y,z}$. The width of the dust ring and the concentration of the dust into a ring are larger for the small Stokes number. The concentration strongly depends on the turbulence strength (shown later in Figs. \ref{fig:Sigma_sum_m0.100} and \ref{fig:Sigma_sum_m0.300}, see also Sect. \ref{sec:Dependence on turbulence strength}).}

\AK{Similar to the flow-shear regime, the advective flux dominates in the distant region far from the planet, $|x|\gg1$,  (\Figref{fig:dust_ring_only}c). The outward (inward) diffusive flux occurs on the right (left) side across the peak of the dust surface density (\Figref{fig:dust_ring_only}c). Inside the planetary orbit, the advective flux is not disturbed because the influence of the outflow toward \editage{the inside of} the planetary orbit is weak. Thus, the dust gap \rev{does} not form in \editage{a} steady\revsec{-}state.}


\subsubsection{Dependence on planetary mass}
\editage{Figures \ref{fig:Sigma_sum_m0.030}--\ref{fig:Sigma_sum_m0.300} (see also Figs. \ref{fig:Contour_min} and \ref{fig:Contour_max}) show the dependence of dust surface density influenced by the planet-induced gas flow on the planetary mass.} We found the following trends: (1) \AK{apparent dust} ring and (or) gap formation by the planet-induced gas flow occurs when
\begin{align}
    m\geq0.1.
\end{align} (2) \AK{Dust} gap formation is susceptible to occur for a higher-mass planet. 

\editage{Regarding} the former trend, recall that the width of the forbidden region increases with the planetary mass, except when the Stokes number is extremely small (${\rm St}\lesssim10^{-3}$; \Figref{fig:<w>}). When $m=0.03$, the smallest planetary mass \AK{considered} in this study, the forbidden region does \rev{exist}, but its width is small ($\langle w_{\rm f}\rangle_{z}\lesssim0.2$; \Figref{fig:<w>}a). \AK{The influence of the planet-induced gas flow on the radial drift of dust is weak. The turbulence diffusion can smooth the dust surface density,} except when the turbulence strength is weak ($\alpha_{\rm diff}\lesssim10^{-5}$; \revsec{Figs. \ref{fig:Sigma_sum_m0.030}g--i}).


\AK{The latter trend} is related to the response of the topologies of the planet-induced gas flow to an increase in the Mach number. \editage{A larger $\mathcal{M}_{\rm hw}$ is needed for a higher-mass planet} to change the \revtir{gas} flow regime from the flow-shear to the flow-headwind regime \AK{(\Equref{eq:flow transition mass})}. Within a range of the Mach numbers considered in this study, when $m=0.3$, the influence of the outflow toward inside the planetary orbit remains and the dust gap forms. \editage{We found the following empirical relation between the planetary mass and the Mach number based on our results.} The dust gap forms when \revsec{(\Figref{fig:Contour_min}):}
\AK{
\begin{align}
    m/\mathcal{M}_{\rm hw}\gtrsim3.
\end{align}
}

\subsubsection{Dependence on turbulence strength}\label{sec:Dependence on turbulence strength}
\revsec{Figures \ref{fig:Contour_min} and \ref{fig:Contour_max}} show the dependence of dust surface density influenced by the planet-induced gas flow on the turbulence strength, $\alpha_{\rm diff}$. In common with \revsec{Figs. \ref{fig:Contour_min} and \ref{fig:Contour_max}}, the influence of the planet-induced gas flow on the dust surface density is stronger for \editage{a} small $\alpha_{\rm diff}$\editage{,} \AK{\editage{in which case} the dust \editage{settles} in the midplane of the disk, and the velocity fluctuations caused by the planet-induced gas flow are large. This is because the outflow \editage{significantly influences} the radial drift of dust near the midplane (\Figref{fig:forbidden_region_m0.300_Mhw0.010}). Then the large surface density gradient needs to achieve a steady\revsec{-}state, which results in a significant accumulation or depletion of dust.}

Within a range of the turbulence strengths considered in this study, the \AK{apparent} dust ring and gap structures are found only when $\alpha_{\rm diff}\leq10^{-4}$ \revsec{(Figs. \ref{fig:Contour_min} and \ref{fig:Contour_max})}. When $\alpha_{\rm diff}$ is very small, $\alpha_{\rm diff}=10^{-5}$, we found an excessive accumulation of dust outside the planetary orbit in some cases \revsec{(Figs. \ref{fig:Contour_max}g--i)}. We \editage{have discussed} an excessive dust concentration outside the planetary orbit in \AK{Sects. \ref{sec:Dust growth} and \ref{sec:Caveats}}. 
\iffigure
\begin{figure}
    \centering
    \includegraphics[width=\linewidth]{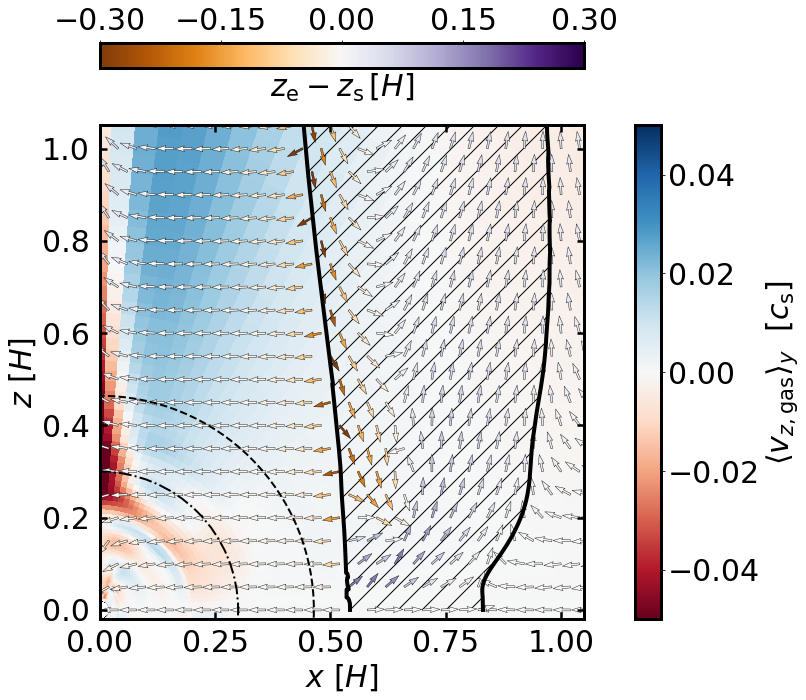}
    \caption{\revtir{Same as \Figref{fig:forbidden_region_m0.300_Mhw0.010}a, but \editage{the} color contour from blue to red represents the $z$-component of the gas flow velocity averaged in the $y$-direction within the calculation domain of hydrodynamical simulation around a planet, $\langle v_{{z},\,{\rm gas}}\rangle_y$. Color contour of the arrows from purple to orange represents the difference in the $z$-coordinates between the ending and starting points of \revfif{orbital calculations}.}}
    \label{fig:delta_z_m0.300_Mhw0.010}
\end{figure}
\fi

\revtir{We assumed that a planet did not perturb the gas surface density (Sect. \ref{sec:Simulation-of-dust-surface-density}), and found that the influences of the planet-induced gas flow on the dust surface density were more pronounced for the small Stokes number (Figs. \ref{fig:Contour_min} and \ref{fig:Contour_max}). However, one might expect that the spatial distribution of dust with the very small Stokes number (${\rm St}\ll1$) would match that of the gas.}

\revtir{\editage{We} consider an endmember case where ${\rm St}=10^{-4}$ and assume that the planet-induced gas flow is in the flow-shear regime. First, we only focus on the dust motion at the midplane region of the disk. Even when ${\rm St}=10^{-4}$, the voids in the dust distribution are created by the outflow of the gas (e.g., \Figref{fig:streamlines-trajectories}b)\editage{,} because the outflow streamlines originated from the horseshoe streamlines at high altitudes \citep[][]{Fung:2015}. \revfif{The 3D structure of the gas flow field is shown in Appendix \ref{sec:Hydrodynamical regimes of the planet-induced gas flow}.}}

\revtir{Next, we consider the vertical distribution of dust. \editage{The transient horseshoe gas flow transports the dust particles at high altitudes to the midplane region} (orange arrows in \Figref{fig:delta_z_m0.300_Mhw0.010}).\footnote{The dust is stirred up by the meridional gas flow, but its magnitude is small (\Figref{fig:delta_z_m0.300_Mhw0.010}). Recent two-fluid (dust + gas) 3D global simulations also suggest that the meridional gas flow induced by a low-mass planet does not effectively stir dust particles \citep{krapp20213d}. \revfif{We discuss the upward motion of dust in Appendix \ref{sec:Dust motion within the forbidden region}.}} After transportation, the dust trajectories at the midplane region should match those shown in \Figref{fig:streamlines-trajectories}b. Thus, the voids remain in the dust distribution.}

\revtir{The downward motion of dust particles is identified even at $z\sim H$ (\Figref{fig:delta_z_m0.300_Mhw0.010}). When the dust scale height is smaller than \editage{the disk scale height}, $H_{\rm d}<H$, dust-poor gas is supplied to the forbidden region. Then the surface density of dust deviates from that of the gas even when ${\rm St}\ll1$. Thus, \revfif{from \Equref{eq:dust scale height}}, our model for the dust substructure formation without the perturbation \editage{of} the gas surface density would require the following condition: $\alpha_{\rm diff}<{\rm St}\ll1$.}


\section{Discussion}    \label{sec:discussion}

\subsection{Comparison to previous studies}
\editage{Many authors have investigated the planet-driven dust ring and gap formation in a disk.} One of the major differences between the conventional models and our model is that \AK{we focus on non-gas-gap-opening, low-mass planets\editage{. Consequently}, the proposed mechanism} works well for the particles with small Stokes numbers.

\AK{\rev{Gas-gap-opening planets} \editage{have long been} a plausible candidate to explain the dust rings and gaps in disks.} A giant planet carves a density gap in a gas disk \citep[e.g.,][]{lin1986tidal}, and then the particles are trapped at the edge of the gas gap. The depth of the dust gap increases with the Stokes number \citep{zhu2012dust,weber2018characterizing}. When particles are tightly coupled with the gas, they can pass through the gas gap due to the turbulent diffusion or the viscous accretion flow \citep{rice2006dust,zhu2012dust,pinilla2016tunnel,weber2018characterizing,Bitsch:2018,drazkowska2019including}.

\AK{Several studies have shown that a low-mass planet that \rev{does} not carve a significant gas gap in a disk can form dust substructures. \revtir{\cite{paardekooper2006dust} found that $\sim15\,M_{\oplus}$ planet perturbs the dust surface density and then forms the dust gap in the distribution of the dust particles larger than $150\,\mu$m. From the analytical point of view, \cite{Muto:2009} showed that $\sim2\,M_{\oplus}$ can open the dust gap when ${\rm St}\gtrsim0.1$ in the absence of the global pressure gradient.} The axisymmetric dust depletion can be generated in an inviscid disk due to steepening of nonlinear density waves \editage{launched} by $\sim8\,M_{\oplus}$ planet when ${\rm St}\gtrsim2\times10^{-3}$ \citep{zhu2014particle}. The annular enhancements in the dust disk \editage{are} caused by the pressure perturbation by a low-mass embedded planet ($\sim15\,M_{\oplus}$) when ${\rm St}\gtrsim2\times10^{-2}$ \citep{rosotti2016minimum}. The depth of the dust gap or the concentration of the dust into a ring increases with the Stokes number \citep{zhu2014particle,rosotti2016minimum}. Dust gap forms due to the tidal torque from a planet ($\sim15\,M_{\oplus}$) when the Stokes number is large, ${\rm St}\gtrsim1$ \citep{dipierro2016two,dipierro2017opening}.} \rev{These previous studies did not consider the influences of the radial outward gas flow induced by an embedded planet, which is the mechanism we study in this paper.}

\editage{Compared} with the \AK{studies mentioned above}, our \editage{study reveals} the opposite trend: the dust ring and gap are likely to form when the Stokes number is small. \AK{The properties of the dust ring and gap, such as width, depth, and amplitude, are more pronounced for the small Stokes number \revsec{(Figs. \ref{fig:Sigma_sum_m0.030}--\ref{fig:Sigma_sum_m0.300})}}\editage{, because} smaller dust particles are more sensitive to the gas flow. \reveig{Our local approach cannot explicitly handle the global effects that investigated in the previous studies (e.g., shallow gas-gap-opening by a low-mass planet), but a local model is needed to reveal the influence of the outflow of the gas.}

\reveig{Here we consider how our model can coexist with the other dust substructure formation mechanisms. In an inviscid disk, low-mass planets could open shallow gas gaps through nonlinear density wave steepening \citep{goodman2001planetary,rafikov2002planet,zhu2014particle,dong2017multiple,dong2018multiple}. Linear waves excited by a low-mass planet steepen to shocks after they propagate a distance \citep{goodman2001planetary}:
\begin{align}
    x_{\rm sh}\approx0.93\,\Biggl(\frac{\gamma+1}{12/5}m\Biggr)^{-2/5}H,\label{eq:linear wave}
\end{align}
where $\gamma=1.4$ is the adiabatic index. Equation (\ref{eq:linear wave}) gives $x_{\rm sh}\sim1.5$--$3.8\,[H]$ within a range of the planetary masses considered in this study, $m=0.03$--$0.3$. These values are larger than the widths of the outflow of the gas, $|x|\lesssim0.5\,[H]$ (e.g., \Figref{fig:streamlines_m0.1}). We would expect that the large dust particles accumulate at $|x|\gtrsim x_{\rm sh}$ due to the shallow gas-gap-opening. Small particles that are tightly coupled with the gas, on the other hand, can diffuse or pass through the gas gap due to the turbulence\revnin{, the viscous accretion flow, or both} \citep[e.g.,][]{rice2006dust}. These small dust are trapped by the outflow of the gas. Thus, when we consider the shallow gas-gap-opening in an inviscid disk, we would expect that the locations of the dust rings and the widths of the dust gaps in the populations of the large and small dust grains do not match each other. The dust ring in the large dust population may appear $\sim1\,[H]$ farther away than that in the small dust population. The width of the dust gap in the large dust population may be wider than that in the small dust.   
}

\subsection{Turbulence in a protoplanetary disk}\label{sec:Turbulence in a protoplanetary disk}
Protoplanetary disks are \AK{thought to be turbulent, though the strength is highly uncertain}. One of the possible origins of the disk turbulence is the magnetorotational instability, which produces strong turbulent intensities of $\alpha_{\rm diff}\sim10^{3}$--$10^{-2}$ \citep[MRI,][]{Balbus:1991}. However, Ohmic diffusion would suppress the MRI, which creates an MRI-inactive region \citep[dead-zone;][]{Gammie:1996}. The dead-zone lies between a few times $0.1$ au and about a few tens of au in a disk, in which the $\alpha_{\rm diff}$ value should be smaller, $\alpha_{\rm diff}\lesssim10^{-4}$ \citep{Gammie:1996}. 

Even if the ionization degree is low, the turbulence \AK{can be} generated through hydrodynamic instabilities, such as vertical shear instability \citep[VSI, e.g.,][]{Urpin:1998,Nelson:2013}. \AK{\editage{The strength} is likely to be lower}. The VSI can develop in the outer region of the disk, $>10$ au, and generate turbulence of $\alpha_{\rm diff}\sim10^{-4}$ \citep[][and references therein]{Malygin:2017}. However, the VSI-driven turbulence may be suppressed by the toroidal magnetic fields in a disk \citep{cui2020global,cui2021vertical}, or dust evolution \citep{fukuhara2021effects}. When the dust grows to $\sim0.1$--$1$mm at $>10$ au (corresponding to ${\rm St}\sim10^{-4}$--$10^{-3}$ in our disk model; \Figref{fig:physical-quantities}c), the VSI may be stabilized completely \citep{fukuhara2021effects}. 

Recent observations suggest low levels of turbulence, \revfif{$\alpha_{\rm diff}\lesssim10^{-5}$--$10^{-3}$}, \editage{from} dust settling \revfif{\citep[][]{pinte2015dust,stephens2017alma,dullemond2018disk,Villenave2022ahighlysettled}}, or direct molecular emission line measurements \AK{\citep[][]{flaherty2015weak,flaherty2017three,flaherty2018turbulence}}. The above findings support \AK{the mechanism we propose\editage{d} in this study as the origin of observed dust rings and gaps,} where \rev{the} weak turbulence condition is required \revsec{(Figs. \ref{fig:Contour_min} and \ref{fig:Contour_max})}. 

\subsection{Size of dust in a protoplanetary disk}\label{sec:Size of dust in a protoplanetary disk}
The planet-induced gas flow \editage{that is} induced by a low-mass planet \editage{strongly influences} on the distribution of aerodynamically small dust, ${\rm St}\sim10^{-4}$--\rev{$10^{-2}$}. Such a small Stokes number corresponds to the dust particles of $\sim0.1$--\rev{$10$} mm in the outer region of the disk, $>10$ au, \AK{for \rev{the steady viscous-accretion disk model} assumed in this study} (\Figref{fig:physical-quantities}c).

\AK{The size of the dust particles in disks can be constrained observationally and theoretically.} Recent observations constrain the maximum dust size in the HL Tau disk: $\sim0.1$ mm \citep[polarization observations at millimeter wavelengths;][]{Kataoka:2016} or $\sim1$ mm \citep[spectral energy distribution (SED) of the millimeter emission;][]{carrasco2019radial}. \editage{Dust coagulation models can also constrain the size of the dust in a disk}. \cite{Okuzumi:2019} has shown that the dust particles covered with the \ce{H2O}-ice mantle can grow up to $\sim50$ mm between the \ce{H2O} and \ce{CO2} snow lines. Inside the \ce{H2O} snow line, where the dust grains exist as nonsticky silicate grains, the maximum dust size \AK{in their model} is $\sim1$ mm. Outside the \ce{CO2} snow line, since the \ce{CO2} ice is as nonsticky as silicate grains \citep{Musiolik:2016a,Musiolik:2016b}, the upper limit of the size of the dust \AK{in their model} is $\sim0.1$--$1$ mm. \AK{The above findings support the mechanism we propose in this study as the origin of observed dust rings and gaps, where small-sized particles are required.}

\subsection{Implications for planet formation}\label{sec:Implications for planet formation}
\subsubsection{Pebble accretion}\label{sec:pebble accretion}
Dust particles with ${\rm St}\sim10^{-3}\text{--}10^0$ can be referred to as pebbles. When the Stokes number is small (${\rm St}\lesssim10^{-2}$), pebble accretion is suppressed significantly by the planet-induced gas flow \citep{Popovas:2018a,Kuwahara:2020a,Kuwahara:2020b,okamura2021growth}. In our previous studies, assuming the uniform \rev{surface density of pebbles}, we calculated the accretion probability of pebbles \citep{Kuwahara:2020a,Kuwahara:2020b}\editage{; however,} \AK{the effect of the planet-induced gas flow on the dust surface density \editage{was} not  considered in these studies}. \AK{In this section, assuming the typical disk model, where the $\mathcal{M}_{\rm hw}$ increases with the orbital radius (\Equref{eq:Mach number}), we consider the accretion rate of pebbles onto a growing protoplanet. The planet-induced gas flow tends to be the flow-shear (flow-headwind) regime in the inner (outer) region of the disk for a fixed dimensionless planetary mass, $m$ (\Equref{eq:flow transition mass}).}

\editage{The dust gap forms around the planetary orbit when the planet-induced gas flow is in the flow-shear regime} (Sect. \ref{sec:Dust ring-gap formation by planet-induced gas flow: flow-shear regime}). \AK{Under the influence of the planet-induced gas flow, the pebbles coming from a narrow window between the horseshoe and the shear regions can accrete onto the planet \citep[\Figref{fig:streamlines-trajectories}b;][]{Kuwahara:2020a}. The accretion cross\editage{-}section of pebbles lies within the dust gap formed by the planet-induced gas flow, because the outflow deflects the trajectories of pebbles outside the horseshoe region (\Figref{fig:streamlines-trajectories}b). Thus, the pebble accretion rate becomes lower than \editage{that estimated} in \cite{Kuwahara:2020a}, where the uniform \rev{pebble} surface density was assumed.} 

\AK{Here we consider the growth of the protoplanet with $m=0.3$ at $1$ au ($M_{\rm pl}\simeq2\,M_{\oplus}$ and $\mathcal{M}_{\rm hw}\simeq0.03$; Figs. \ref{fig:physical-quantities}a and b). Following \cite{Kuwahara:2020a}, we assume $\alpha_{\rm diff}=10^{-4}$ and ${\rm St}=10^{-3}$ (see Sect. 4.4.3 of \cite{Kuwahara:2020a} for a detailed description). Assuming uniform surface density of pebbles, the accretion rate of pebbles is $\dot{M}_{\rm acc}\sim3\,M_{\oplus}\text{/Myr}$ \citep{Kuwahara:2020a,okamura2021growth}. \rev{When the effects of planet-induce gas flow on the dust distribution is considered,} the pebble surface density \rev{would be} reduced by a factor of $\sim2$ due to the dust gap \revsec{(\Figref{fig:Contour_min}e)}. Thus, the effective accretion rate could be reduced to $\sim1.5\,M_{\oplus}\text{/Myr}$. The growth timescale can be estimated as $M_{\rm p}/\dot{M}_{\rm acc}\sim1\,\text{Myr}$\editage{, which} could be much longer when the turbulence strength \AK{or the Stokes number} is smaller ($\alpha_{\rm diff}<10^{-4}$ \AK{or ${\rm St}<10^{-3}$}).}



\editage{Only a dust ring is likely to be formed when the planet-induced gas flow is in the flow headwind regime} (e.g., Figs. \ref{fig:Sigma_sum_m0.100}e and f). \AK{In this case, the accretion rate of pebbles could increase due to an increase in the pebble surface density near the planetary orbit.}

\AK{We continue to consider the growth of the protoplanet, but we consider $m=0.1$ (\rev{$M_{\rm pl}\simeq3\,M_{\oplus}$ at $10$ au}). We assume $\mathcal{M}_{\rm hw}=0.1$, which means that we focus on the outer region of the disk, $\gtrsim10$ au, where the planet-induced gas flow tends to be the flow-headwind regime. Following \cite{Kuwahara:2020a}, we assume $\alpha_{\rm diff}=10^{-4}$ and ${\rm St}=3\times10^{-3}$. In the flow-headwind regime, the accretion probability of pebbles does not differ significantly from that obtained in the unperturbed flow except when the Stokes number is very small \citep[${\rm St}\lesssim10^{-3}$;][]{Kuwahara:2020b}. The accretion rate of pebbles can be estimated as $\sim10\,M_{\oplus}$/Myr for the uniform surface density of pebbles \citep{Ormel:2017-pebble,Liu:2018,Ormel:2018,Kuwahara:2020a}. The pebble surface density \rev{would be} increased by a factor of $\sim2$ due to the dust ring \revsec{(\Figref{fig:Contour_max}f)}. Thus, the effective accretion rate could be increased to $\sim20\,M_{\oplus}\text{/Myr}$. In the outer region of the disk, the VSI-driven turbulence could be suppressed \citep[][see also Sect. \ref{sec:Turbulence in a protoplanetary disk}]{fukuhara2021effects}. This leads to \editage{a} significant accumulation of pebbles \revsec{(\Figref{fig:Contour_max}i)} and could result in more efficient growth of the planet via pebble accretion.}

\subsubsection{\rev{Implications for the origins of the architecture\revsec{s} of planetary system\revsec{s}}}
\rev{Based on the discussion in Sect. \ref{sec:pebble accretion}, we consider a formation scenario of planetary systems. In our previous studies, assuming  uniform surface density of pebbles, we found that only the suppression of pebble accretion due to the planet-induced gas flow occurs in the late stage of planet formation ($m\gtrsim0.03\text{--}0.1$), more specifically, in the inner region of the disk \citep[$\lesssim1$ au;][]{Kuwahara:2020a,Kuwahara:2020b}.}

\rev{Our \editage{current results} further supports the above discussions\editage{, which} suggests that the growth of the protoplanets via pebble accretion is suppressed in the inner region of the disk ($\lesssim1$ au). When the mass of the protoplanets reaches $m\gtrsim0.1$ ($\gtrsim0.7\,M_{\oplus}$ at $1$ au), the subsequent growth of the \revsec{protoplanets} is highly suppressed by the reduction of the accretion probability of pebbles and pebble surface density due to the planet-induced gas flow. \revsec{These protoplanets can avoid runaway gas accretion within the lifetime of the gas disk, which results in} the formation of rocky terrestrial planets. The small rocky protoplanets may also experience giant impacts. These events lead to the formation of super-Earths.}

\rev{\editage{In contrast}, the efficient growth of the protoplanets via pebble accretion would be achieved in the outer region of the disk ($\gtrsim10$ au). Thus, we expect that the planets could exceed the critical core mass within the typical lifetime of the disk,\editage{leading to formation of giant planets.}}

\rev{Our scenario may \editage{help explain} the distribution of exoplanets (the dominance of super-Earths at $<1$ au; \citealt{Fressin:2013,Weiss:2014} and a possible peak in the occurrence of gas giants at $\sim2\text{--}3$ au \citealt{Johnson:2010,Fernandes:2019}), as well as the architecture of the Solar System: rocky planets in the inner regions, gas giants in the middle, and icy giants in the outer regions.}

\rev{The reduction of the pebble isolation mass in the inner region of the inviscid disk may cause the dichotomy that is apparent between the inner super-Earths and outer gas giants \citep{Fung:2018}. However, the small pebbles (${\rm St}\lesssim10^{-3}$) may not be trapped at the local pressure maxima and thus, they continue to contribute to the growth of the planet \citep{Bitsch:2018}. Since the smaller pebbles are more sensitive to the gas flow, the planet-induced gas flow has the potential to explain the dichotomy without the pressure maxima.}

\subsubsection{\AK{Dust growth and planetesimal formation}} \label{sec:Dust growth}
From Figs. \ref{fig:Sigma_sum_m0.100} and \ref{fig:Sigma_sum_m0.300}, we found an excessive accumulation of small dust outside the planetary orbit in some cases. An increase in the dust-to-gas ratio may trigger planetesimal formation via direct growth and streaming instability \citep[SI;][]{Kretke:2007,Youdin:2005,Youdin-Johansen:2007,Johansen:2007}. The growth rate of SI depends on the size of dust, which has the maximum value when ${\rm St}\sim10^{-1}\text{--}10^{0}$ \citep{chen2020efficient,umurhan2020streaming}. Thus, SI does not work for small dust grains. 

In this study, we did not consider dust growth, \rev{but} the accumulation of dust leads to subsequent dust growth due to an increase in the collision frequency of dust, and then the Stokes number increases. Assuming perfect sticking via collisions without collisional fragmentation \rev{and bouncing}, the growth timescale can be described by $t_{\rm grow}\sim(\reveig{\Sigma_{\rm g}}/\Sigma_{\rm d})\Omega_{\rm K}^{-1}$ \citep{takeuchi2005attenuation}. Comparing the growth timescale to the drift timescale, when $t_{\rm grow}\ll t_{\rm drift}$, the dust grains enter the "growth-dominated phase" \citep{taki2021new}. \cite{taki2021new} derived the Stokes number in the equilibrium state as follows:
\begin{align}
    {\rm St}_{\rm eq}\approx\frac{\sqrt{\pi}}{8\eta}\frac{\Sigma_{\rm d}}{\reveig{\Sigma_{\rm g}}},
\end{align}
where $\eta$ is given by \citep{Ida:2016}:
\tiny
\begin{empheq}[left={\eta\simeq\empheqlbrace}]{alignat=2}
&\displaystyle0.93\times10^{-3}\,\Biggl(\frac{M_{\ast}}{1\,M_{\odot}}\Biggr)^{-7/10}\Biggl(\frac{\alpha_{\rm acc}}{10^{-3}}\Biggr)^{-1/5}\Biggl(\frac{\dot{M}_{\ast}}{10^{-8}\,M_{\odot}/\text{yr}}\Biggr)^{2/5}\Biggl(\frac{r}{1\,\text{au}}\Biggr)^{2/20},\label{eq:eta viscous}\\
&\displaystyle0.80\times10^{-3}\,\Biggl(\frac{L_{\ast}}{1\,L_{\odot}}\Biggr)^{2/7}\Biggl(\frac{M_{\ast}}{1\,M_{\odot}}\Biggr)^{-8/7}\Biggl(\frac{r}{1\,\text{au}}\Biggr)^{4/7}.\label{eq:eta irradiation}
\end{empheq}
\normalsize
Equations (\ref{eq:eta viscous}) and (\ref{eq:eta irradiation}) are valid in the region where viscous heating and stellar irradiation dominate, respectively. \rev{The dimensionless value of} $\eta$ has on the order of $10^{-3}$ and does not change significantly across the entire region of the disk ($\lesssim100$ au). Thus, ${\rm St}_{\rm eq}$ strongly depends on the local dust-to-gas ratio. When ${\rm St}_{\rm eq}\gtrsim1$, the dust particles rapidly grow to planetesimal through the positive feedback between the reduced radial drift and the accelerated collisional growth \citep{taki2021new}.

\editage{From Figs. \ref{fig:Sigma_sum_m0.100} and \ref{fig:Sigma_sum_m0.300},} the above condition is easily satisfied outside the planetary orbit under weak turbulence. As a result, the dust particles grow rapidly and begin to decouple from the gas. The planetesimals could form outside the planetary orbit, \rev{where dust accumulate\revsec{s} due to the planet-induced \revtir{gas} flow.} 

\subsection{Implications for \revsec{protoplanetary disk} observations}
The gas flow induced by low-mass planets can be considered  one of the possible origins of the observed dust \editage{substructures} in disks. The characteristics of the dust ring and gap formed by the planet-induced gas flow are summarized as follows:
\begin{enumerate}
    \item \rev{The planet-induced gas flow forms the ring and (or) gap in aerodynamically small dust grain \editage{distribution}, ${\rm St}\lesssim10^{-2}$.}
    \item The gas flow induced by a low-mass embedded planet can generate the dust ring and (or) gap \rev{in a radial \revtir{extent} of  $\sim1\text{--}10\,[H]$ around the planetary orbit} without creating a gas gap or pressure bump.
\end{enumerate}

Our model for the dust ring and gap formation by low-mass planets can be tested when the following conditions are satisfied: (1) the dust substructures appear only for the distribution of small dust grains ($\lesssim1$ cm). (2) There is no correlation between the spatial distribution of gas and dust in a disk. This information can be obtained from observations of dust emission and scattering in different wavelengths to constrain the dust surface densities for different sizes and of spectral lines which trace the gas surface density 

\editage{In the following sections, we have discussed how our model can be distinguished from the other dust substructure formation mechanisms.}

\subsubsection{Correlation between the spatial distribution of gas and dust}
\AK{Our model focuses on the low-mass planets \editage{that} neither open gas \editage{gaps} nor generate pressure \editage{bumps}. Thus, correlations between the spatial distribution of gas and dust would not be observed.}

\AK{When the dust substructures form due to the \rev{gas-gap-opening} planet, \editage{however}, there would be a correlation between the spatial distribution of gas and dust in disks, such as transition disks where the inner disks are depleted in dust relative to the outer disk \citep{francis2020dust}. In many transition disks, correlations between emission lines of gas and continuum rings are observed at the outer edge of the cavity,  which suggests the existence of a massive planet within the cavity \citep[][]{van2016resolved,van2018new,dong2017sizes,boehler2017close,fedele2017alma}.}

\subsubsection{Kinematic features in a protoplanetary disk}
\rev{When a massive planet is embedded in a disk, kinematic features appear in the three velocity components of gas in \editage{the} upper disk layers due to the large-scale meridional \revtir{gas} flow \citep{Morbidelli:2014,szulagyi2014accretion,fung2016gap}. The vertical flows toward the midplane with speeds of $\sim0.1c_{\rm s}$ at a region $\sim2\text{--}4\,[H]$ above the midplane of the disk around HD 163296 \editage{have} been detected by observations of \ce{^12CO} emissions, which can be interpreted as a sign of the large-scale meridional \revtir{gas} flow that driven by Jupiter-mass planets \citep{teague2019meridional}.
}

\rev{Low-mass planets that do not carve a gap can induce the gas flow such as the small-scale meridional \revtir{gas} flow \citep{Fung:2015} or the subsonic polar inflow and radial outflow \citep[e.g.,][]{Ormel:2015b}. These flows occur within the typical scale of the planetary atmosphere \citep[$\lesssim\min(R_{\rm Bondi},R_{\rm Hill})$;][]{Fung:2015,Lambrechts:2017,Kuwahara:2019,Bethune:2019,Fung:2019,krapp20213d}. Thus, as long as we focus on the low-mass planets, the kinematic features of gas that can be traceable in upper layers of disks by molecular emissions, e.g., \ce{^12CO} ($J=2\text{--}1$), would not be expected to \revsec{be observable}. 
}

\rev{Meanwhile, the vertical location of the molecular emission surface depends on the molecular isotopologues and species. For instance, since the relative abundances of \ce{CO} isotopologues are different, the emission surface approaches the midplane of the disk in the order of \ce{ ^12CO}, \ce{^13CO}, \ce{C^18O}, and \ce{C^17O}. Other molecules, such as \ce{HCN} and \ce{C_2H},
can be used as tracers. The \ce{C_2H} lines come from a lower height than \ce{CO} isotopologue \citep{law2021molecules,bosman2021molecules}. When \editage{observations with multiple chemical species constrain the velocity structure of the disk} \citep{yu2021mapping} and the kinematic features of the gas can only be seen in the region close to the midplane ($z\lesssim1\,[H]$) at the orbital location of the dust substructure, such features of gas and dust support our model.}

\subsection{\AK{Caveats}}\label{sec:Caveats}
\subsubsection{\revsec{Spatial distribution of dust in the global domain}}\label{sec:Caveats: Spatial distribution of dust in the global domain}
\rev{The distribution of dust particles outside the local computational domain of orbital calculation is nontrivial. The planet-induced gas flow deflects the trajectories of particles and creates the voids in the $x$-$y$ plane \citep[][see also \Figref{fig:streamlines-trajectories}b]{Kuwahara:2020a,Kuwahara:2020b}. Outside the local domain, however, these voids might be filled by the drift of particles in the $x$-direction, \revsec{and} turbulent diffusion. Our local simulations could not handle the void closing. In \editage{estimating dust drift velocity, we assumed that the effects of the planet-induced gas flow are} negligible outside the domain of the orbital calculations (namely, the void is closed outside the domain). This assumption is a conservative approach to give a minimum effect of the planet-induced gas flow on dust drift. In other words, if a planet can open a gap in our 1D model under this assumption, it means that gap could form even if we treat the global dust motion explicitly.}

\subsubsection{\revsec{Steady-state solution}}
\AK{In this study, we only provide the steady\revsec{-}state solution of the 1D advection-diffusion equation (\Equref{eq:advection-diffusion equation}). The time scale to reach the steady\revsec{-}state depends on an inward advective flux of dust from the outer region of the disk. The drift speed of dust is \editage{low} when the Stokes number is small. Thus, the advective flux of dust is small for a fixed initial dust surface density.  It might take a long time to reach a steady-state.}

\reveig{The time scale to reach the steady-state can be estimated by the time scale \revnin{for} the dust concentration \revnin{in} a ring. The dust accumulates outside the planetary orbit with \revnin{the timescale $t_{\rm accum}$:}
\begin{align}
    t_{\rm accum}\sim\frac{M_{\rm ring}}{\dot{M}_{\rm d}},
\end{align}
where $M_{\rm ring}$ is the total mass of the mass of the ring and $\dot{M}_{\rm d}$ is the radial inward mass flux of dust. The ring mass can be described by
\begin{align}
    M_{\rm ring}=\tilde{M}_{\rm ring}H^2\Sigma_{\rm d,0},
\end{align}
where $\tilde{M}_{\rm ring}$ is the dimensionless mass of the ring that is obtained from our numerical results (Figs. \ref{fig:Sigma_sum_m0.030}--\ref{fig:Sigma_sum_m0.300}). The initial dust surface density can be described by
\begin{align}
    \Sigma_{\rm d,0}\simeq\frac{\dot{M}_{\rm d}}{2\pi r|v_{\rm drift}|}=\frac{\dot{M}_{\rm d}}{2\pi r^2\Omega_{\rm K}h^2}\Biggl(\frac{1+{\rm St^2}}{\rm St}\Biggr)\Biggl(\frac{\mathrm{d}\ln \revnin{p}}{\mathrm{d}\ln r}\Biggr)^{-1}. \label{eq:sigma_d_0}
\end{align}
From \Equref{eq:gas surface density}, the gas surface density in a steady accretion disk can be rewritten as \citep{Ida:2016}
\footnotesize
\begin{align}
    \Sigma_{\rm g}\simeq\frac{10^{-5}}{6\pi^2}\Biggl(\frac{\dot{M}_{\ast}}{10^{-8}\,M_{\odot}\text{/yr}}\Biggr)\Biggl(\frac{M_\ast}{1\,M_{\odot}}\Biggr)^{-1/2}\Biggl(\frac{r}{1\,\text{au}}\Biggr)^{1/2}\Biggl(\frac{\alpha_{\rm acc}}{10^{-3}}\Biggr)^{-1}h^{-2}\,[M_{\odot}/\text{au}^2].\label{eq:sigma_g}
\end{align}
\normalsize
From Eqs. (\ref{eq:sigma_d_0}) and (\ref{eq:sigma_g}), $\Sigma_{\rm d,0}$ can be expressed as
\footnotesize
\begin{align}
    \Sigma_{\rm d,0}\simeq\frac{10^{-5}}{\pi^2}\Biggl(\frac{1+{\rm St}^2}{\rm St}\Biggr)\Biggl(\frac{\dot{M}_{\rm d}}{10^{2}\,M_{\oplus}\text{/Myr}}\Biggr)\Biggl(\frac{M_\ast}{1\,M_{\odot}}\Biggr)^{-1/2}\Biggl(\frac{r}{1\,\text{au}}\Biggr)^{1/2}h^{-2}\,[M_{\oplus}/\text{au}^2].
\end{align}
\normalsize
Thus, the time scale for \revnin{the} dust accumulation can be estimated by
\begin{align}
    t_{\rm accum}\simeq\frac{10^{-7}\times\tilde{M}_{\rm ring}}{\pi^2}\Biggl(\frac{1+{\rm St}^2}{\rm St}\Biggr)\Biggl(\frac{M_\ast}{1\,M_{\odot}}\Biggr)^{-1/2}\Biggl(\frac{r}{1\,\text{au}}\Biggr)^{5/2}\text{Myr}.\label{eq:t_acc}
\end{align}
Here we consider the planet with $m=0.1$, the Stokes number ${\rm St}=10^{-4}$, and the turbulent parameter $\alpha_{\rm diff}=10^{-4}$. When $\mathcal{M}_{\rm hw}=0.03\,(\sim1\,\text{au}$; \Figref{fig:Sigma_sum_m0.100}e), \Equref{eq:t_acc} gives $t_{\rm accum}\sim6000$ yr. When $\mathcal{M}_{\rm hw}=0.1\,(\sim50\,\text{au}$; \Figref{fig:Sigma_sum_m0.100}f), \Equref{eq:t_acc} gives $t_{\rm accum}\sim133$ Myr.}

\revnin{From these estimations, in the inner region of the disk ($\lesssim1$ au), the dust surface density can reach the steady-state within the typical disk lifetime of a few Myr \citep{2001ApJ...553L.153H}, but it seems difficult to reach the steady-state in the outer region ($\gtrsim10 $ au). We plan to investigate the time evolution of the dust surface density influenced by the planet-induced gas flow in the second paper of this series. In reality, the time evolution of dust surface density is determined by complex processes such as dust growth and fragmentation. These additional processes should be included in further studies.}

\section{Conclusions} \label{sec:conclusion}

\iffigure
\begin{figure}
    \centering
    \includegraphics[width=\linewidth]{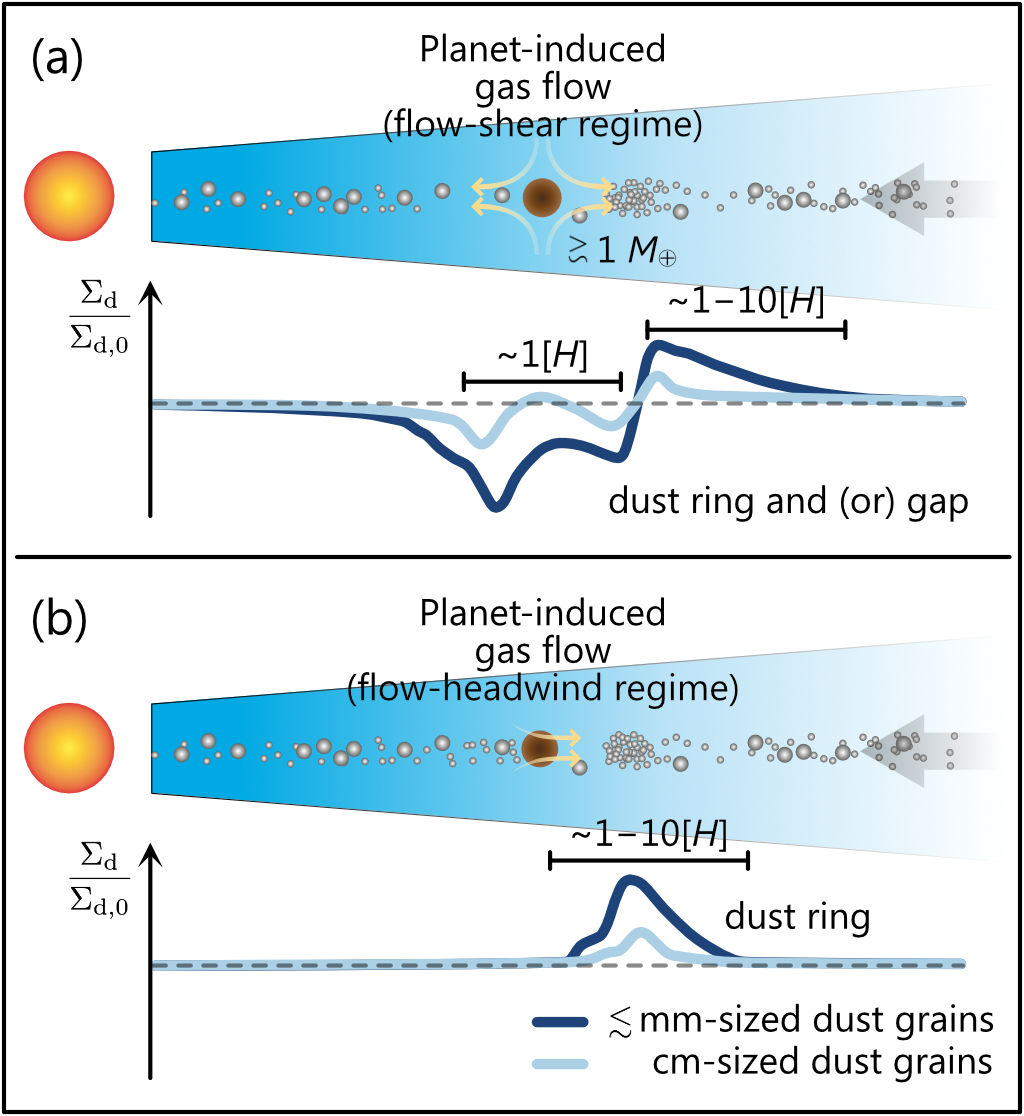}
    \caption{Schematic illustrations of our model for the dust ring and gap formation. The planet-induced gas flow is in the flow-shear regime (\textit{panel a}) and in the flow-headwind regime (\textit{panel b}) to form dust substructures.}
    \label{fig:summary}
\end{figure}
\fi

\revtir{We investigated the influence of the gas flow induced by low-mass, non-gas-gap-opening planets on the spatial distribution of dust in protoplanetary disks. First, we performed 3D hydrodynamical simulations, which resolve the local \revtir{gas} flow in the region\editage{,} including the Hill and Bondi spheres of the planet. We then calculated the trajectories of dust influenced by the planet-induced gas flow. We used the obtained hydrodynamical simulation data to calculate the gas drag force acting on the dust particle. Finally, we computed the steady-state dust surface density by incorporating the influences of the planet-induced gas flow into the 1D advection-diffusion equation. We summarize our main findings as follows (\revfif{\Figref{fig:summary}}).}

\begin{enumerate}
\item When ${\rm St}\lesssim3\times10^{-2}$ ($\lesssim1$ cm at $\sim1\text{--}10$ au), a forbidden region exists outside the planetary orbit due to the outflow of the gas, where \editage{the planet-induced gas flow inhibits the radial drift of dust particles.} The width of the forbidden region is wider for small\revsec{er} Stokes number\revsec{s}.
\item Under weak turbulence ($\alpha_{\rm diff}\lesssim10^{-4}$), the gas flow induced by a non-gas-gap-opening planet with $m\gtrsim0.1$ ($\gtrsim0.7\,M_{\oplus}$ at $1$ au, or $\gtrsim3.3\,M_{\oplus}$ at $10$ au) generates the dust accumulation (\revsec{the} dust ring) and depletion (\revsec{the} dust gap) with the radial \revtir{extent} of $\sim1\text{--}10\,[H]$ around the planetary orbit in the distribution of small dust grains, ${\rm St}\lesssim10^{-2}$ ($\lesssim1$ cm at $\sim1\text{--}10$ au). When the outflow occurs dominantly both outside and inside the planetary orbit ($m/\mathcal{M}_{\rm hw}\gtrsim3$), the dust ring and (or) gap form(s). When the outflow occurs dominantly outside the planetary orbit ($m/\mathcal{M}_{\rm hw}\lesssim3$), only the dust ring forms outside the planetary orbit in the steady\revsec{-}state. 
\item The properties of the dust ring and gap, such as width, depth, and amplitude, are more pronounced for the small Stokes number\editage{,} because smaller dust particles are more sensitive to the gas flow.
\end{enumerate}

\rev{The dust ring and gap formation due to the planet-induced gas flow \editage{significantly impacts} the processes of planet formation. The planetary growth via pebble accretion would be inefficient due to dust depletion around the planetary orbit. This is susceptible to occur in the inner region of the disk ($\lesssim1$ au). \editage{In contrast, the dust accumulation could result in efficient growth of the planet via pebble accretion}, and may trigger planetesimal formation via rapid growth of the dust particles. This is susceptible to occur in the outer region of the disk ($\gtrsim1\text{--}10$ au). \revsec{Our result may be helpful to explain the dichotomy between the inner super-Earths and outer gas giants.}}

The gas flow induced by low-mass, non-gas-gap-opening planets can be considered one of the possible origins of the observed dust substructures in disks. 
\revtir{When (1) the dust ring and gap appear only for the distribution of small dust grains ($\lesssim1$ cm), and (2) there is no correlation between the spatial distribution of gas and dust in a disk, such dust substructures may indicate the existence of a low-mass planet embedded in a protoplanetary disk ($\gtrsim1\,M_{\oplus}$). }


\begin{acknowledgements}
\reveig{We would like to thank an anonymous referee for constructive comments.} \rev{We thank Athena++ developers. This study has greatly benefited from fruitful discussions with Satoshi Okuzumi, Ayaka Okuya, and Yuya Fukuhara. Numerical computations were in part carried out on Cray XC50 at the Center for Computational Astrophysics at the National Astronomical Observatory of Japan. \revsec{This work} was supported by JSPS KAKENHI Grant number 20J20681, \reveig{20K04051}, \revten{and 21H04514}.}
\end{acknowledgements}




\newpage
\appendix
\def\thesection{A}
\setcounter{equation}{0}
\def\theequation{A.\arabic{equation}}
\setcounter{figure}{0}
\def\thefigure{A.\arabic{figure}}


\section{\revfor{Gas drag regimes}}\label{sec:Gas drag regimes}
The gas drag force is divided into two regimes, the Epstein and the Stokes regimes, depending on the relationship between the size of the particle and mean free path of the gas (Eqs. (\ref{eq:Epstein-regime}) and (\ref{eq:Stokes-regime}))\editage{, which} \AK{differ in terms of the dependence on the gas density. In this study, we assumed that the gas drag force is independent of the gas density, which is strictly valid only in the Stokes regime \rev{but \editage{not} in the Epstein regime}. However,} the region of particular interest in this study is typically beyond the Bondi region, where the gas density is almost \AK{constant and identical to the initial (unperturbed) value}. We confirmed that the choice of the gas drag regime does not affect the results \AK{and thus\editage{,} the results can be applicable for dust particles both in Epstein and Stokes regimes.}
\def\thesection{B}
\setcounter{equation}{0}
\def\theequation{B.\arabic{equation}}
\setcounter{figure}{0}
\def\thefigure{B.\arabic{figure}}

\section{\revfor{Maximum impact parameter of accreted \revfif{particles}}}\label{sec:Maximum impact parameter of accreted particle}
The maximum impact parameter of accreted particles in the unperturbed flow, $b$, is expressed by \citep{Ormel:2010,Lambrechts:2012,Guillot:2014,Ida:2016,Sato:2016}:
\begin{align}
    b = b_0\exp\left[-\left(\frac{\rm St}{2}\right)^{0.65}\right],\label{eq:impact parameter}
\end{align}
where
\begin{align}
    b_0\simeq\min\left(2\sqrt{\frac{m{\rm St}}{\mathcal{M}_{\rm hw}}},2{\rm St}^{1/3}R_{\rm Hill}\right).
\end{align}
\def\thesection{C}
\setcounter{equation}{0}
\def\theequation{C.\arabic{equation}}
\setcounter{figure}{0}
\def\thefigure{C.\arabic{figure}}

\section{\reveig{Density waves excited by an embedded planet}}\label{sec:Density waves excited by an embedded planet}

\iffigure
\begin{figure}
    \centering
    \includegraphics[width=\linewidth]{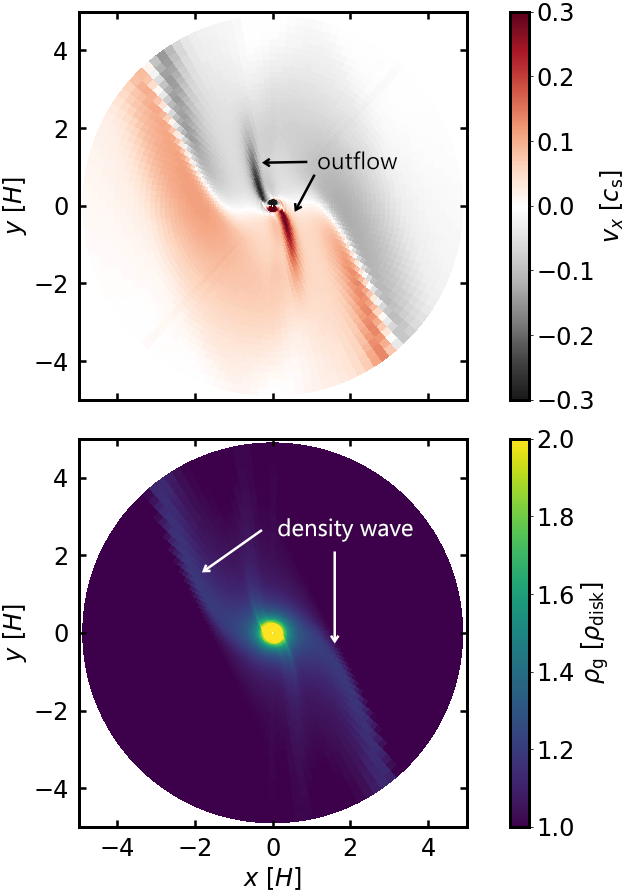}
    \caption{\reveig{Flow structure at the midplane of the disk around a planet obtained from \texttt{m03-hw001}. The color contours represent the flow speed in the $x$-direction (\textit{top}) and the gas density (\textit{bottom}).}}
    \label{fig:density waves}
\end{figure}
\fi

\reveig{An embedded planet excites the density waves as shown in \Figref{fig:density waves}. Since the density waves extend in the azimuthal direction, they are interrupted at the edge of the computational domain of our hydrodynamical simulations, $|x|\sim3\,[H]$. Figure \ref{fig:density waves} also shows that the outflow of the gas \revnin{appears} closer to the planetary orbit than the density waves.}

\def\thesection{D}
\setcounter{equation}{0}
\def\theequation{D.\arabic{equation}}
\setcounter{figure}{0}
\def\thefigure{D.\arabic{figure}}

\section{\revfor{Temperature and aspect ratio of the steady accretion disk}}\label{sec:Temperature and aspect ratio of the steady accretion disk}
The disk midplane temperature is given by $T_{\rm disk}=\max(T_{\rm vis},\,T_{\rm irr})$, where $T_{\rm vis}$ and $T_{\rm irr}$ are temperatures determined by viscous heating and stellar irradiation \citep{garaud2007effect,oka2011evolution,Ida:2016},
\begin{align}
    T_{\rm vis}\simeq&200\,\Biggl(\frac{M_{\ast}}{1\,M_{\odot}}\Biggr)^{3/10}\Biggl(\frac{\alpha_{\rm acc}}{10^{-3}}\Biggr)^{-1/5}\Biggl(\frac{\dot{M}_{\ast}}{10^{-8}\,M_{\odot}/\text{yr}}\Biggr)^{2/5}\Biggl(\frac{r}{1\,\text{au}}\Biggr)^{-9/10}\,{\rm K},\label{eq:T vis}\\
    T_{\rm irr}\simeq&150\,\Biggl(\frac{L_{\ast}}{1\,L_{\odot}}\Biggr)^{2/7}\Biggl(\frac{M_{\ast}}{1\,M_{\odot}}\Biggr)^{-1/7}\Biggl(\frac{r}{1\,\text{au}}\Biggr)^{-3/7}\,{\rm K},\label{eq:T irr}
\end{align}
where $\dot{M}_{\ast}$ is the accretion rate of the disk and $L_{\ast}$ is the stellar luminosity. In this study, we assume a solar-mass host star, $M_{\ast}=1\,M_{\odot}$, the solar luminosity, $L_{\ast}=1\,L_{\odot}$, \revfor{and} the typical accretion rate of classical T Tauri stars, $\dot{M}_{\ast}=10^{-8}\,M_{\odot}/\text{yr}$.

The aspect ratio of the disk is given by
\begin{align}
    h\equiv \frac{H}{r}=\max(h_{\rm g,vis},\,h_{\rm g,irr}),\label{eq:aspect-ratio}
\end{align}
where $h_{\rm g,vis}$ and $h_{\rm g,irr}$ are aspect ratios given by \citep{Ida:2016}:
\begin{align}
    h_{\rm g,vis}\simeq&0.027\,\Biggl(\frac{M_{\ast}}{1\,M_{\odot}}\Biggr)^{-7/20}\Biggl(\frac{\alpha_{\rm acc}}{10^{-3}}\Biggr)^{-1/10}\Biggl(\frac{\dot{M}_{\ast}}{10^{-8}\,M_{\odot}/\text{yr}}\Biggr)^{1/5}\Biggl(\frac{r}{1\,\text{au}}\Biggr)^{1/20},\label{eq:h vis}\\
    h_{\rm g,irr}\simeq&0.024\,\Biggl(\frac{L_{\ast}}{1\,L_{\odot}}\Biggr)^{1/7}\Biggl(\frac{M_{\ast}}{1\,M_{\odot}}\Biggr)^{-4/7}\Biggl(\frac{r}{1\,\text{au}}\Biggr)^{2/7}.\label{eq:h irr}
\end{align}

\def\thesection{E}
\setcounter{equation}{0}
\def\theequation{E.\arabic{equation}}
\setcounter{figure}{0}
\def\thefigure{E.\arabic{figure}}

\section{\revfor{Hydrodynamical regimes of the planet-induced gas flow}}\label{sec:Hydrodynamical regimes of the planet-induced gas flow}

\iffigure
\begin{figure*}
    \centering
    \includegraphics[width=\linewidth]{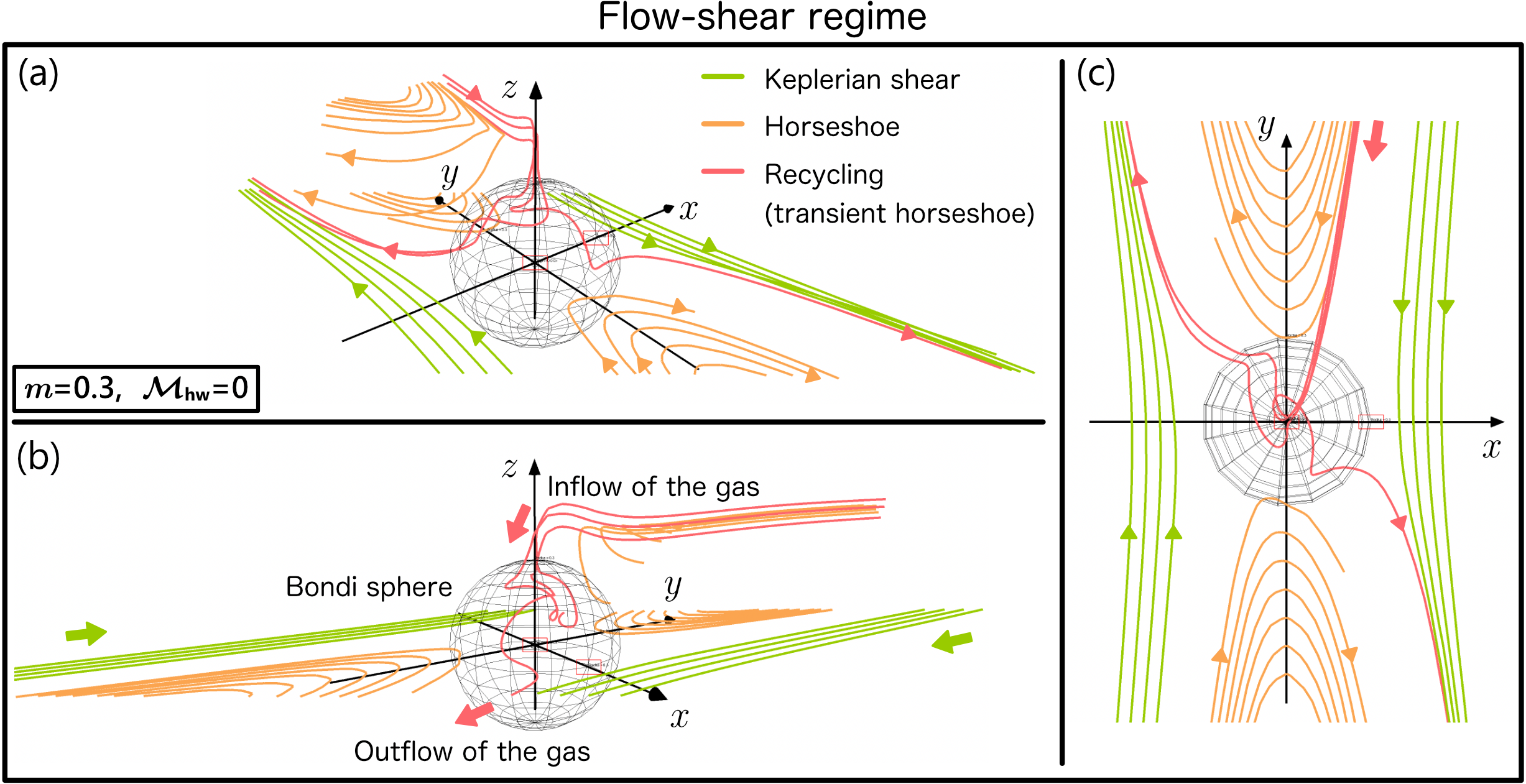}
    \caption{Streamlines of 3D planet-induced gas flow around the planet with $m=0.3$ and $\mathcal{M}_{\rm hw}=0$. The planet-induced gas flow is in the flow-shear regime. The result obtained in \cite{Kuwahara:2020a}. The green, orange, and red solid lines are the Keplerian shear, the horseshoe streamlines, and the recycling streamlines \citep{Ormel:2015b,Kuwahara:2019} (or transient horseshoe streamlines \citep{Fung:2015}), respectively. The sphere is the Bondi sphere of the planet whose size is 0.3 [H]. The arrows represent the direction of the gas flow. \textit{Panels a and b:} perspective views of the flow field. \textit{Panel c}: $x$–$y$ plane viewed from the +$z$-direction.}
    \label{fig:flow-shear}
\end{figure*}
\fi

\iffigure
\begin{figure*}
    \centering
    \includegraphics[width=\linewidth]{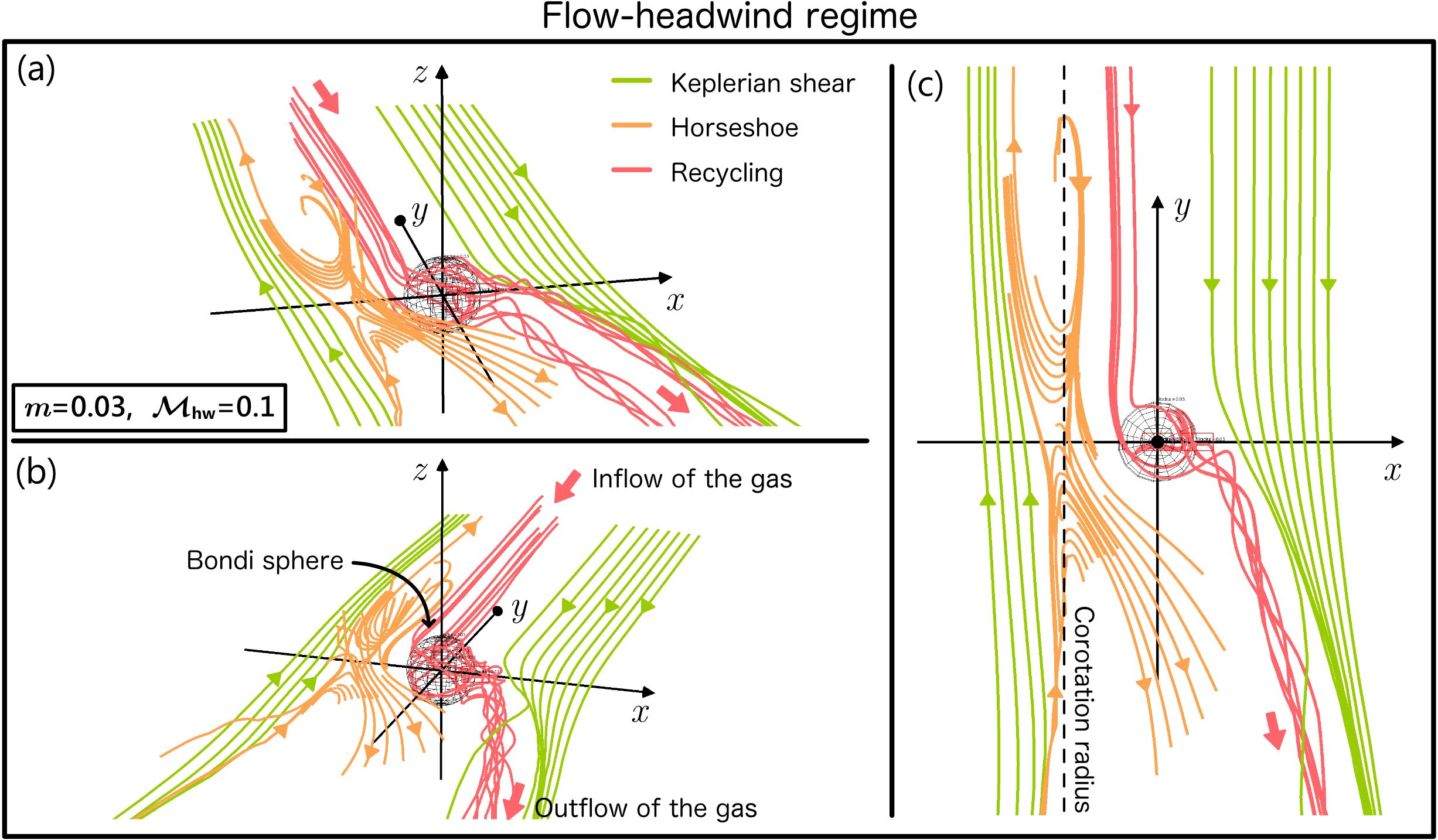}
    \caption{Same as in \Figref{fig:flow-shear}, but with $m=0.03$ and $\mathcal{M}_{\rm hw}=0.1$. The planet-induced gas flow is in the flow-headwind regime. The sphere is the Bondi sphere of the planet whose size is 0.03 [H]. The vertical dashed line in \textit{panel c} corresponds to the $x$-coordinate of the position of the corotation radius for the gas, $x_{\rm g,cor}=-2/3\mathcal{M}_{\rm hw}$.}
    \label{fig:flow-headwind}
\end{figure*}
\fi

\revfor{The gas flow around an embedded planet is perturbed by the planet, which forms the characteristic 3D structure of the flow field \citep{Ormel:2015b,Fung:2015,Fung:2019,Cimerman:2017,Lambrechts:2017,Kurokawa:2018,Kuwahara:2019,Bethune:2019,moldenhauer2021steady}. The planet-induced gas flow is divided into two regimes, the flow-shear and flow-headwind regimes, depending on the relationship between the dimensionless planetary mass and the Mach number of the headwind of the gas \citep[\Equref{eq:flow transition mass};][]{Kuwahara:2020b}.} %

\revfor{The fundamental features of the flow field are as follows: (1) gas from the disk enters the Bondi or Hill sphere (inflow) and exits through the midplane region of the disk (outflow; the red lines of Figs. \ref{fig:flow-shear} and \ref{fig:flow-headwind}). (2) The horseshoe flow exists in the anterior-posterior direction of the orbital path of the planet (the orange lines of Figs. \ref{fig:flow-shear} and \ref{fig:flow-headwind}). The horseshoe streamlines have a columnar structure in the vertical direction (Figs. \ref{fig:flow-shear}a and b). (3) The Keplerian shear flow extends inside and outside the orbit of the planet (the green lines of Figs. \ref{fig:flow-shear} and \ref{fig:flow-headwind}).}

\def\thesection{F}
\setcounter{equation}{0}
\def\theequation{F.\arabic{equation}}
\setcounter{figure}{0}
\def\thefigure{F.\arabic{figure}}

\section{\revfor{Dust motion within the forbidden region}}\label{sec:Dust motion within the forbidden region}

\iffigure
\begin{figure}
    \centering
    \includegraphics[width=\linewidth]{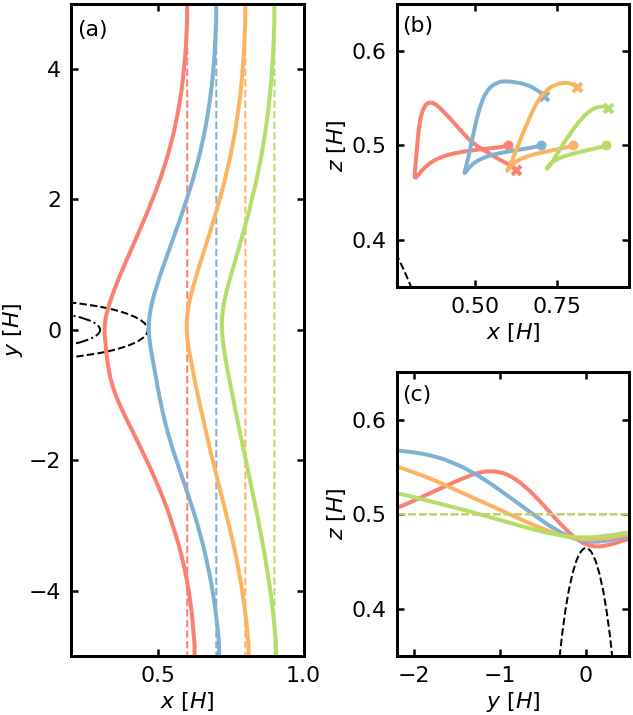}
    \caption{\AK{Trajectories of dust particles with ${\rm St}=10^{-4}$ \revtir{in the planet-induced gas flow (solid lines) and in the unperturbed sub-Keplerian shear flow (dashed lines)}. We use\rev{d} the result obtained from \texttt{m03-hw001} to calculate dust trajectories. We set $z_{\rm s}=0.5$ for the orbital calculations. Different colors correspond to different $x_{\rm s}$. The filled circles and cross symbols in \textit{panel b} correspond to $x_{\rm s}$ and $x_{\rm e}$, respectively. The dotted and dashed circles are the Bondi and Hill radii of the planet, respectively. \textit{Panel a}: $x$-$y$ plane viewed from the $+z$ direction. \textit{Panel b}: $x$-$z$ plane viewed from the $-y$ direction. \textit{Panel c}: $y$-$z$ plane viewed from the $+x$ direction. This figure should be compared to \Figref{fig:forbidden_region_m0.300_Mhw0.010}a.}}
    \label{fig:orbit_m0.3_hw0.01_St0.0001_z=0.000}
\end{figure}
\fi

\revsev{We first describe the motion of dust particles in the $x$-direction.} When the Stokes number is small \rev{(${\rm St}=10^{-4}$)}, the forbidden region lies in a wide range of \Figref{fig:forbidden_region_m0.300_Mhw0.010}a. \AK{\rev{Counterintuitively}, we found that the forbidden region exists even above the outflow \rev{whose vertical scale is $\sim0.5R_{\rm Bondi}$. To understand this phenomenon, we show} the trajectories of dust particles with ${\rm St}=10^{-4}$ at high altitudes \rev{in \Figref{fig:orbit_m0.3_hw0.01_St0.0001_z=0.000}}. \revtir{Figure \ref{fig:orbit_m0.3_hw0.01_St0.0001_z=0.000} compares the dust trajectories in the planet-induced gas flow (solid lines) with those in the unperturbed gas flow (dashed lines).} In the distant region far from the planet, $y>-4$, the orbital radii of dust in the planet-induced gas flow are larger than those in the unperturbed gas flow (\Figref{fig:orbit_m0.3_hw0.01_St0.0001_z=0.000}a). Figure \ref{fig:orbit_m0.3_hw0.01_St0.0001_z=0.000}b shows $x_{\rm e}>x_{\rm s}$ in the planet-induced gas flow. This is caused by the gas flow structure at high altitudes. \revsev{Note that} the changes in the orbital radii of dust should be carefully considered. Since our hydrodynamical simulations were local, we did not trace the gas flow structure in the full azimuthal direction. Thus, our results may need to be compared with the results obtained by global hydrodynamical simulations \revfif{in a future study}.

When the Stokes number is large, the width of the forbidden region decreases significantly above the outflow (\Figref{fig:forbidden_region_m0.300_Mhw0.010}b). \rev{The width of the forbidden region \revsec{for} ${\rm St}=10^{-2}$ (\Figref{fig:forbidden_region_m0.300_Mhw0.010}b) is narrower than that \revsec{for} ${\rm St}=10^{-4}$ (\Figref{fig:forbidden_region_m0.300_Mhw0.010}a).} \rev{When ${\rm St}=3\times10^{-2}$ (\Figref{fig:forbidden_region_m0.300_Mhw0.010}c)}, the width of the forbidden region is quite narrow ($\lesssim0.1\,[H]$). \AK{This is because, when the Stokes number is large, the dust particles are less affected by the gas flow.}

\revsev{Next, we describe the motion of dust particles in the $z$-direction.} The dust particles \revtir{influenced by the planet-induced gas flow} ascend  when they pass through near the planet location (\revtir{solid lines in} Figs. \ref{fig:orbit_m0.3_hw0.01_St0.0001_z=0.000}b and c).} \revtir{The upward motion of dust particles is caused by the meridional gas flow induced by the planet. \revsev{In the case of low-mass planets, the meridional gas flow occurs as follows:} the outflow streamlines connect to the horseshoe streamlines \citep[called "transient horseshoe flow";][]{Fung:2015}. In the flow-shear regime, gas flows in at high latitudes of the gravitational sphere of the planet \editage{at} its horseshoe turn. The transient flow is vertically compressed, then the upward gas flow occurs due to the decompression of the gas \citep[see e.g., Fig.11 of][]{Fung:2015}.}

\end{document}